\documentclass[a4paper,reprint,
superscriptaddress,
 amsmath,amssymb,
 aps,
 prc,
longbibliography,
dvipsnames,
accepted=2023-04-05,
twocolumn]{quantumarticle}
\pdfoutput=1
\usepackage[utf8]{inputenc}

\usepackage[main=english,german,ngerman]{babel}

\usepackage{xcolor}
\usepackage{graphicx}
\usepackage{dcolumn}
\usepackage{bm}
\usepackage{siunitx}
\usepackage{dsfont}
\usepackage[ruled]{algorithm2e}
\usepackage{amsthm}

\usepackage{ulem}

\usepackage[numbers]{natbib}

\usepackage{mathtools}
\usepackage{ragged2e}

\usepackage{hyperref}

\usepackage{bm}

\usepackage{dsfont}
\usepackage{subcaption}

\usepackage{mathtools }

\usepackage{physics}
\usepackage{bbold}
\usepackage{float}
\usepackage{hyphenat}

\usepackage[shortlabels]{enumitem}
\newtheorem{lemma}{Lemma}
\newtheorem{definition}{Definition}
\newtheorem{theorem}{Theorem}
\newtheorem{corollary}{Corollary}

\usepackage[utf8]{inputenc}
\usepackage[toc,page]{appendix}

\newcommand{\update}[1]{#1}

\begin{document}

\title{Importance sampling for stochastic quantum simulations}

\author{Oriel~Kiss}
\orcid{0000-0001-7461-3342}
\email{oriel.kiss@cern.ch}
\affiliation{European Organization for Nuclear Research (CERN), Geneva 1211, Switzerland}
\affiliation{Department of Nuclear and Particle Physics, University of Geneva, Geneva 1211, Switzerland}
\author{Michele~Grossi}
\orcid{0000-0003-1718-1314}
\affiliation{European Organization for Nuclear Research (CERN), Geneva 1211, Switzerland}
 
\author{Alessandro~Roggero}
\orcid{0000-0002-8334-1120}
\affiliation{Physics Department, University of Trento, Via Sommarive 14, I-38123 Trento, Italy}
\affiliation{INFN-TIFPA Trento Institute of Fundamental Physics and Applications, Trento, Italy}

\begin{abstract}
Simulating many\hyp body quantum systems is a promising task for quantum computers. However, the depth of most algorithms, such as product formulas, scales with the number of terms in the Hamiltonian, and can therefore be challenging to implement on near\hyp term, as well as \update{early} fault\hyp tolerant quantum devices. An efficient solution is given by the stochastic compilation protocol known as qDrift, which builds random product formulas by sampling from the Hamiltonian according to the coefficients. In this work, we unify the qDrift protocol with importance sampling, allowing us to sample from arbitrary probability distributions, while controlling both the bias, as well as the statistical fluctuations. We show that the simulation cost can be reduced while achieving the same accuracy, by considering the individual simulation cost during the sampling stage.
 
Moreover, we incorporate recent work on composite channel and compute rigorous bounds on the bias and variance, showing how to choose the number of samples, experiments, and time steps for a given target accuracy. These results lead to a more efficient implementation of the qDrift protocol, both with and without the use of composite channels. Theoretical results are confirmed by numerical simulations performed on a lattice nuclear effective field theory.

\end{abstract}

\maketitle

\section{Introduction}

The simulation of quantum systems is \update{arguably} one of the most promising applications for quantum computers \cite{feynman}. Hence, the exponential scaling of the Hilbert space, or the infamous sign problem in Monte Carlo techniques \cite{sign_Troyer}, makes it a notoriously difficult task for classical devices. On the other hand, the resource requirements for quantum simulations are only subject to polynomial growth in many practical circumstances, as in the simulation of local Hamiltonians~\cite{Lloyd_1996}, \update{and more particularly in spin chains \cite{science_quantum_sim_cost}.} Hence, quantum computers offer a natural paradigm for Hamiltonian simulations, with numerous applications in nuclear~\cite{Dumitrescu_2018,roggero2019, PRC_Kiss} and condensed matter physics \cite{Hofstetter_2018,quantum_simulations_XXZ,lmg_grossi,PRB_LMG,Monaco_PRB}, quantum field theory \cite{QS_QFT_Preskill, QS_Schwinger_Lougovski,Klco_2022} and quantum chemistry \cite{Su2021nearlytight, Ouyang2020compilation, Martinez_partitioning,Li6_batteries_arrazola,1stq_Babbush}. For instance, quantum simulations have been applied to the computation of energy levels via quantum phase estimation \cite{QPE-Lloyd}, chemical reaction rates predictions \cite{reaction}, correlation functions \cite{two_point_roggero,Hofstetter_2018,quantum_simulations_XXZ,nature_grossi}, neutrino oscillations \cite{PRD_neutrino,neutrino_simulation_amitrano} and scattering experiments \cite{neutrio_nucleus_roggero, du2021,Illa_2022}. 

Given a Hamiltonian $H$ written as the sum of $L$ multi\hyp qubit operators, typically expressed as Pauli strings, the solution of the time\hyp independent Schr\"odinger equation is obtained through the exponentiation of the Hamiltonian in question. One of the most popular and straightforward techniques to compute this matrix exponential is given by product formulas, such as Trotterization and higher order Trotter\hyp Suzuki decomposition \cite{Suzuki, SUZUKI1990319, Wiebe_2010_trotter, PhysRevX_high_trotter}, due to their simplicity and high performance in practice, \update{which is usually much better than the worst case analytical error bounds~\cite{random_input_Child, science_quantum_sim_cost}.}

However, one main drawback of product formulas is their relative high cost to accuracy ratio. Hence, their gate count increases proportionally to the number of summands $L$ in the Hamiltonian. Even if the asymptotic scaling is favorable, the pre\hyp factor might be significant enough~\cite{Troyer_gate_count, T_gate_Troyer,neutrio_nucleus_roggero} to create problems in practice. Hence, deep circuits usually exceed the coherence time of NISQ devices, and would also be problematic for early fault\hyp tolerant quantum hardware as well. \update{Following~\cite{Preskill2018quantumcomputingin} we define (N)ISQ devices as (noisy) intermediate scale quantum devices, composed of a few hundred qubits, and equipped with error correction (error mitigation) protocols that can correct up to a limited amount of error. Hence, both NISQ and ISQ devices are limited in size and depth, making it a priority to improve the complexity of current quantum algorithms.} We refer to~\cite{RevModPhys.86.153} and~\cite{simulation_tacchino,quantum_simulations_lanes} for informative reviews, where the former is theoretical, and the latter focus on practical applications on NISQ devices. 

Randomization has proven to be an important tool to improve the accuracy and efficiency of product formulas. \textcite{Childs2019fasterquantum} achieved better gate\hyp complexity by randomizing the ordering of the terms in the Hamiltonian, \textcite{Faehrmann2022randomizingmulti} increased the order of the product formula by averaging over different time slices, \update{while \textcite{random_QPE} proposed a randomized quantum phase estimation procedure independent of $L$.} More recently, \textcite{doubling_random_berry} doubled the order of product formulas by introducing random corrections. Hence, stochasticity can turn coherent errors into incoherent ones, making the error scaling behave as a random walk \cite{PhysRevA_random_knee,PhysRevA_random_wallman,PhysRevLett_Poulin,Tran_2021}. Even if those improvements are considerable, none of these methods really address the scaling with $L$, which can be a large pre\hyp factor for many relevant situations.

For this reason, \textcite{QDrift} introduced qDrift, a protocol to build product formulas by sampling over the coefficients, whose length does not depend specifically on $L$, but on the square of the simulation time $t^2$ and on the spectral norm of the Hamiltonian $\lambda$. \textcite{QDRift_caltech} improved the error bound on the bias and showed that one experiment of the qDrift protocol converges exponentially fast towards its expectation value, while \textcite{Ouyang2020compilation} combined the advantages of qDrift and first\hyp order Trotter formula to simulate the Hamiltonian through sparsification. \update{Finally, \textcite{qSWIFT}  proposed a higher order qDrift protocol, known as qSwift, by adding correction terms to the standard qDrift approach.}

The main contributions of this paper are the generalization of the qDrift protocol for arbitrary sampling distribution and the expansion of the results from Ref.~\cite{QDRift_caltech} to multiple qDrift executions and composite channels with multiples Trotter steps. Sampling from arbitrary distributions has numerous benefits, such as allowing for a direct reduction of the actual simulation's cost in terms of native gates, or expanding qDrift to situations where it might be difficult to sample directly from the coefficients. For instance, we propose an alternative sampling distribution, which decreases the simulation cost, such as the total CNOT count. Moreover, this paper gives a rigorous understanding of the behavior of qDrift and composite channels with multiple experiments. We show that qDrift can be efficiently parallelized on multiples devices, and we give a rigorous formula for choosing the number of qDrift samples $N$, experiments $M$, and Trotter steps $r$, for a given accuracy $\epsilon$ and simulation time $t$. \update{We hope this paper to be a starting point to build qDrift protocols tailored to certain applications and computing devices.}

Even if we restrict ourselves to time\hyp independent problems,  we note that the present work can be directly applied to time\hyp dependent situations through the continuous qDrift \cite{Berry2020timedependent} protocol. We note that alternative and more refined techniques exist, using extra ancillary qubits and complex gadgets, such as qubitization\update{/quantum signal processing \cite{QSP_Chuang,Low2019hamiltonian}}, and linear combination of unitaries \cite{Child,Black_box_Berry, PhysRevX.8.041015}. They usually offer better asymptotic scaling, but are more challenging to implement in practice and fall outside this paper's scope.

The relevant background and notations are covered in Section~\ref{sec:bg}, with Section~\ref{sec_qdrift} recalling the qDrift protocol. We introduce importance sampling for stochastic quantum simulations in Section~\ref{sec:isqdrift}, while rigorous bounds on the bias, variance, and fluctuation are shown in Section~\ref{sec_bounds}. Section~\ref{sec_composite} unifies the importance sampled qDrift with composite channels, mixing Trotter and qDrift product formulas. Section~\ref{applications} focuses on practical applications that benefit from this general framework, such as cost reduction and Hamiltonian partitioning. Finally, numerical simulations are performed in Section~\ref{sec_simulation}. Rigorous proofs of the stated theorems can be found in the appendices \ref{app:main} and \ref{app:cost} for the results and applications sections, respectively.

\section{Preliminaries}
\label{sec:bg}
In this section, we introduce the background and notations adopted throughout this paper. We denote by $\|\cdot\|$ the spectral norm, $\|\cdot\|_1$ the trace norm and by 
\begin{align}
\label{diamond_norm}
     \norm{\mathcal{U}- \mathcal{E}}_\diamond & := \max_\rho{\norm{(\mathcal{U}\otimes \mathbb{1}_k - \mathcal{E} \otimes \mathbb{1}_k)\rho}_1 },
\end{align}
the diamond norm distance between two quantum channels $\mathcal{U}$ and $\mathcal{E}$. We note that the maximum is taken over all $(n+k)$\hyp qubit states, where $n$ is the dimension of $\mathcal{E}$ and $k\geq1$.
In the remaining of this paper, we consider a time\hyp independent $n$\hyp qubit Hamiltonian, in the form of a $(2^n,2^n)$ Hermitian matrix with the following decomposition into $L$ summands
\begin{equation}
\label{h}
    H = \sum_{l=1}^L h_l H_l,
\end{equation}
 with $\norm{H_l}=1$ and $h_l>0$. We denote $\lambda = \sum_l h_l$ the norm of the Hamiltonian and $\Lambda \coloneqq \max_{l}(h_l)$.

\subsection{Deterministic Trotter product formulas}
Given a Hamiltonian $H$, the first order Trotter product formula is built by exponentiating all the individual terms as
\begin{equation}
    \label{trotter}
    U(t) = \prod_l e^{-ith_lH_l}.
\end{equation} 
 The usual technique for long\hyp time simulations is to split the time into $r$ fragments and to repeatedly apply $U(t/r)$, which are known as Trotter steps. Analytical work \cite{Childs2019fasterquantum} shows that the Trotter error $\epsilon$ is upper bounded by 
\begin{equation}
    \epsilon \leq \frac{L^2 \Lambda^2 t^2}{2r} e^{\Lambda tL/r}.
\end{equation}
Better error scaling can be achieved by considering higher order product formula, which typically requires a symmetric extension or randomization, leading to deeper circuits, while tighter error bounds with commutator scaling have been found by \textcite{Theory_Trotter}. Despite theirs simplicity, Trotter formulas are performing surprisingly well and often much better than the predicted bounds, making them the default choice in many situations, including early fault tolerant quantum hardware.

Since we will be dealing with quantum channels, it is useful to recall how time evolution is performed in the density matrix formalism. Given a unitary operator $U(t)$ and a density matrix $\rho$, the time evolution is computed as 
\begin{equation}
    \rho(t) = \mathcal{U}(t)[\rho] = U(t)\rho U^\dagger(t),
\end{equation}
where $\mathcal{U}(t)$ is the unitary channel of $U(t)$.
 
\subsection{The qDrift protocol} 
\label{sec_qdrift}
One of the main drawbacks of Trotter\hyp Suzuki decompositions is their gate count. Hence, every term in the Hamiltonian, see Eq.~\eqref{h}, must be simulated sequentially, leading to deep circuits in many relevant use cases. The resulting quantum circuits are therefore heavily affected by noise on NISQ devices and are time\hyp consuming on fault\hyp tolerant hardware, motivating the search for more efficient algorithms. \textcite{QDrift} remarked that the gate count can be significantly reduced in some regimes by considering the relative importance of each term $H_j$ in the Hamiltonian, given by the corresponding coefficient $h_j$. The qDrift protocol builds an approximate channel ${\mathcal{E}}(t;N,M)$, which is randomly constructed by sampling terms from the Hamiltonian according to the magnitude of the coefficients. We call $t$ the simulation time, $N$ the number of qDrift samples, and $M$ the number of qDrift experiments. In practice, a qDrift experiment is sampled from the product distribution $p_N(\bm{j})=\lambda^{-N} \prod_{k=1}^N$ $h_{j_k}$, where each terms is sampled with probability $p(j) = h_j/\lambda$, and $\bm{j} = (j_1, \,j_2, \,...,\, j_N)$ a multi\hyp index, leading to 
\begin{equation}
\label{qdrift_unitary}
    V_{\bm{j}}(t) = \prod_{k=1}^N e^{-i\tau_{j_k} H_{j_k}}\;,
\end{equation}
for appropriately chosen time steps $\tau_{j_k}$.
The qDrift channel is ultimately built as the arithmetic average of the $M$ individual experiments
\begin{equation}
    \label{qdrift_channel_M}
    \mathcal{E}(t;N,M)[\rho] = \frac{1}{M}\sum_{m}^M\left[ V_{\bm{j}^m} \rho V^\dagger_{{\bm{j}^m}}\right],
\end{equation}
as summarised in Algorithm \ref{alg:qdrift}, with the sampling distribution $q(j) = p(j)\coloneqq h_j/\lambda$. We note that the superscript $m$ of the bold multi\hyp index $\bm{j}^m$ refers to the multi\hyp index $\bm{j}$ of the $m$-th experiment, while $j^m_k$ \update{ to its corresponding $k$-th component.} We note that we use the notation $q(j)$ and $q_j$ interchangeably throughout the paper.

In the asymptotic limit of infinite experiments, this will then converge to the following average qDrift channel
\begin{equation}
\begin{split}
\label{eq:qdrift_average}
\overline{\mathcal{E}}(t;N)[\rho] &= \mathbb{E}_{p_N}\left[\mathcal{E}(t;N,1)[\rho]\right]\\
&=\sum_{j_1=1}^L\cdots\sum_{j_N=1}^Lp_N(\bm{j})V_{\bm{j}}(t)\rho V^\dagger_{\bm{j}}(t)\;,
\end{split}
\end{equation}
\update{where $\mathbb{E}_p[f(x)]=\sum_x p(x)f(x)$ is the expectation value of an arbitrary function $f(x)$ and sampling distribution $p(x)$.}
The special case with $N=1$ leads to
\begin{equation}
\label{qdrift_channel}
    \overline{\mathcal{E}}(t;1)[\rho] = \sum_{j=1}^{L}p(j) e^{-i\tau_j H_{j}} \rho e^{i\tau_j H_{j}},
\end{equation}
which we will call the deterministic qDrift channel.

\begin{algorithm}[t]
\caption{\textsc{stochastic quantum \\ {\color{white} empty spacei }simulation}}
\label{alg:qdrift}
\KwIn{Hamiltonian $H=\sum_{l=1}^L h_lH_l$ with interaction strength $\lambda = \sum_l h_l$, $h_l>0$ and $\norm{H_l}=1$, total simulation time $t$, number of samples $N$, number of experiments $M$ and initial state $\rho$.
Probability density distribution $q(j)$.}
$m \gets 1$\;
$\mathcal{E} \gets 0$\;
\While{$m \leq M$}{
    $V \gets \mathbb{1}$\;
    $n \gets 1$\;
    \While{$n \leq N$}{
        sample $j \sim q(j)$\;
        $\tau_j\gets th_j/(q(j)N)$\;
        $V\gets e^{-iH_j\tau_j}V$\;
        $n\gets n+1$\;
    }
    $\mathcal{E}[\rho] \gets \mathcal{E}[\rho] + V\rho V^\dagger$ \;
    $m \gets m+1$\;
}
$\rho \gets  \frac{1}{M}\mathcal{E}[\rho]$\;
\KwOut{final state $\mathcal{E}(t;N,M)[\rho]$}
\end{algorithm}

The strength of each unitary is fixed to a constant $\tau_j :=\tau = t\lambda/N$, which is chosen such that the qDrift channel is equal to the first order Trotter product formula, up to first order in the Taylor expansion,  with an upper bound on the diamond norm given by 
\begin{align}
     \norm{\mathcal{U}(t) - \overline{\mathcal{E}}(t;N)}_\diamond \leq \frac{2\lambda^2t^2}{N}e^{2\lambda t/N}\;,
\end{align}
We note that this bound has been improved~\cite{QDRift_caltech} to 
\begin{align}
\label{eq:qdrift_ebound}
     \norm{\mathcal{U}(t) - \overline{\mathcal{E}}(t;N)}_\diamond \leq \frac{2\lambda^2t^2}{N}.
\end{align}
The deterministic channel can be used to parallelize a Trotter product formula for small time scales, where every $H_l$ is simulated simultaneously and independently on different quantum devices, or by using different qubits of the same device. This scheme essentially trades the circuit's depth with measurements. Expectations values can then be obtained by post\hyp processing. However, since the number of circuits grows exponentially with the number of slicing steps $r$, this application is impractical for long\hyp time scales or large $L$, but could benefit NISQ devices which often fail in this regime due to the noise.

\subsection{Importance sampling} 
\label{sec_is}
Importance sampling is a useful technique to compute expectation values
\begin{equation}
    \mathbb{E}_p[f(x)] = \sum_x p(x)f(x),
\end{equation} when the distribution $p(x)$ is difficult to sample from, or as a way to reduce the variance. Frequently used in Monte Carlo integration, the trick is to sample from an alternative, in some cases considerably easier, distribution $q(x)>0$ and re\hyp weight accordingly 
\begin{equation}
     \mathbb{E}_p[f(x)] = \sum_x q(x)\frac{p(x)}{q(x)}f(x)\equiv \mathbb{E}_q[\omega(x)f(x)],
\end{equation}
with $\omega(x):= p(x)/q(x)$ is the re\hyp weighting factor. We can easily see that this gives us an unbiased estimator of the expectation value of interest, while the variance depends on the choice of $q(x)$. With an adequate choice, the variance can be significantly reduced, leading to less expensive calculations. We guide the reader to \cite{importance_sampling_review} for an informative review on the topic. 

\section{Results}
The first results presented in this section, are the application of importance sampling to the qDrift protocol and the computation of the corresponding bias, variance and fluctuation bounds. The second set of results unifies this framework with composite channels. We will begin by introducing the importance sampled qDrift.
\subsection{Importance sampled qDrift}
\label{sec:isqdrift}
To better understand the paradigm shift, we adopt the Liouvillian representation of a unitary channel $e^{-iHt}$
\begin{equation}
    \mathcal{E}(t)[\rho] = e^{iHt}\rho e^{-iHt} \equiv e^{t\mathcal{L}}(\rho) =\sum_{n=0}^{\infty}\frac{t^n \mathcal{L}^n(\rho)}{n!},
\end{equation}
with 
\begin{equation}
    \mathcal{L}(\rho) = i(H\rho - \rho H) = i [H,\rho].
\end{equation}
We first write the qDrift channel, sampled from an arbitrary distribution $q(j)$ as

\begin{equation}
\begin{split}
\bar{\mathcal{E}}_q(t;1)[\rho]=&\sum_j q(j) e^{-i\tau_jH_j}\rho e^{i\tau_jH_j}\\
\equiv &\sum_j q(j) e^{\tau_j\mathcal{L}_j}(\rho)\\
=&\left(1+\sum_j q(j)\tau_j \mathcal{L}_j+\sum_{n=2}^{\infty}\sum_j q(j)\tau^n_j \mathcal{L}^n_j\right)(\rho)\;.
\end{split}
\end{equation}
We then choose $\tau_j$ so that we match the ideal channel to linear order, that is
\begin{equation}
\begin{split}
\sum_j q(j)\tau_j \mathcal{L}_j &= \frac{t}{N}\sum_j h_j \mathcal{L}_j\\
&= \frac{t\lambda}{N} \sum_j \frac{h_j}{\lambda} \mathcal{L}_j\\
&= \frac{t\lambda}{N} \sum_j p_j \mathcal{L}_j\;.
\end{split}
\end{equation}
Since the bias at the second order in $t/N$ cannot be matched with any choice of $q(j)$, we focus on the part which is linear in time. For ease of notation, we incorporate the constant factor inside the generators as
\begin{equation}
\widetilde{\mathcal{L}_j}=\frac{t\lambda}{N}\mathcal{L}_j\;.
\end{equation}
The expectation value of a qDrift sample is then in the right form and can be written as 
\begin{equation}
\begin{split}
\mathbb{E}_p\left[\widetilde{\mathcal{L}_j}\right]&=\sum_j p_j \widetilde{\mathcal{L}_j}\\
&=\sum_j q_j \frac{p_j}{q_j}\widetilde{\mathcal{L}_j}\\
&=\sum_j q_j \omega_j \widetilde{\mathcal{L}_j}\\
&=\mathbb{E}_q\left[\omega_j\widetilde{\mathcal{L}_j}\right].
\end{split}
\end{equation}
Therefore, the channel can be written as
\begin{equation}
\begin{split}
\bar{\mathcal{E}}_q(t;1)[\rho]&=\sum_j q_j \exp\left(\omega_j \frac{t\lambda}{N}\mathcal{L}_j(\rho)\right)\\
&=\sum_j q_j \exp\left(\frac{t}{N}\frac{h_j}{q_j}\mathcal{L}_j(\rho)\right),\\
\end{split}
\end{equation}
where we used the explicit expression for $p_j$, and find \begin{equation}
\label{eq:tauj}
    \tau_j = \frac{t h_j}{Nq_j}.
\end{equation}
We note that, contrary to the standard qDrift, the strength of each unitary $\tau_j$ is now dependent on $j$. \update{The procedure is summarised in Algorithm \ref{alg:qdrift}, which is a generalisation of the regular qDrift protocol for arbitrary sampling distribution.}

\subsection{Bias, variance and fluctuation bounds}
\label{sec_bounds}
In this section, we compute the bias, concentration and fluctuation bounds for the importance sampled qDrift channel, as a function of the sampling distribution $q(j)$, simulation time $t$, number of samples $N$ and number of experiments $M$. We remark that for $q(j)=p(j)$, we recover the usual bounds, meaning that our framework is a natural extension of the standard qDrift implementation. For the ease of notations, all minimum values of $N$, $M$ and $r$, which are integer numbers per definition, are given in function of $\epsilon,\, t$ and $\lambda$, which may not lead to integer value numbers. Hence, they should be rounded  up to the next integer in practical situations. Borrowing the proof strategy from \cite[Proposition 3.2~]{QDRift_caltech}, we will now provide an upper bound on the error of the bias. We refer to Appendix \ref{app:main} for the complete proofs of the presented results.

\begin{theorem}[Bias error bound]
\label{th_tighter_bound}
Let $\mathcal{U}(t)$ be the unitary channel of a first\hyp order Trotter product formula, $\overline{\mathcal{E}}_q(t;N)$ an average qDrift channel with importance sampling and $\omega(j) = p(j)/q(j)$ the re\hyp weighting factor. The diamond norm distance between these two channels for $N=1$ is then upper bounded by
\begin{equation}
\norm{\mathcal{U}\left(t\right)-\overline{\mathcal{E}}_q\left(t;1\right)}_\diamond \leq t^2\lambda^2 \left(1+\mathbb{E}_p\left[\omega(j) \right]\right),
\end{equation}
leading to the following result 
\begin{equation}
\norm{\mathcal{U}\left(t\right)-\overline{\mathcal{E}}_q(t;N)}_\diamond \leq \frac{t^2\lambda^2}{N} \left(1+\mathbb{E}_p\left[\omega(j) \right]\right)
\end{equation}
for an arbitrary $N$.
\end{theorem}

We are now able to understand that importance sampling can lead to an increase in the number of qDrift samples $N$ at fixed accuracy $\epsilon$. In fact
\begin{equation}
\label{eq_N}
N_q = \left\lceil \frac{t^2\lambda^2}{\epsilon}(1+\mathbb{E}_p[\omega(j)]) \right\rceil \geq \left\lceil \frac{t^2\lambda^2}{\epsilon}(1+1) \right\rceil = N_p,
\end{equation}
implying $N_q\geq N_p$. Therefore, the standard qDrift channel will always requires a smaller number of samples, however, as we will argue in the next section, the total simulation cost can still be reduced by using importance sampling, without sacrificing accuracy, since we can favorise the sampling of cheaper circuits.

Now that we have generalized the error bound on the bias, we need to understand how a finite importance sampled qDrift channel concentrates around its expectation value. Hence, this will provide an estimate of $M$ and $N$, for a given accuracy $\epsilon$ and simulation time $t$.

\begin{theorem}[Concentration bound]
\label{th:3}
Let $\mathcal{E}_q(t;N,M)$  be a finite importance sampled qDrift channel on $n$ qubits and $V_j$ instances of the $NM$ unitaries that make up the channel. Their concentration around their expectation value can then be upper bounded $\forall \epsilon \in [0,4t\lambda]$ as follows
\begin{equation}
\begin{split}
    \text{\normalfont{Pr}}&\left[\norm{\frac{1}{M}\sum_m^M \prod_{k=N}^1 V_{\update{{j}_k^m}} - \mathbb{E}_q\left[V_j\right]^N} \geq \epsilon/2 \right]  \\
   &\leq 2^{n+1} \exp{- \frac{NM\epsilon^2}{11t^2\lambda^2(1+\max_k\omega(k))^2}}\;.
    \end{split}
\end{equation}
In order to guarantee an approximation error $\epsilon/2$ with probability at least $1-\delta$, it is then sufficient to take
\begin{equation}
NM = 11\frac{t^2\lambda^2}{\epsilon^2}\left(1+\max_k\omega(k)\right)^2(n+1)\log\left(\frac{2}{\delta}\right)\;.
\end{equation}
\end{theorem}

This theorem gives us two important pieces of information. First, we learn that we can distribute the resource budget across $M$ and $N$, and that, for fixed accuracy $\epsilon$, the channel converges exponentially fast towards the deterministic qDrift with their product. Moreover, the qDrift channel can be efficiently simulated in parallel, since it concentrates exponentially fast in $M$. This trade-off between circuit's depth for an increase in measurements has some advantages, such as reducing hardware errors on NISQ devices and shorter real\hyp time simulation on fault\hyp tolerant ones. Secondly, we are now aware that the qDrift results present in the literature also hold for any distribution $q(j)$, as long as their ratio is similar, i.e., if $\omega(j) \approx 1$. This enables the design of alternative distributions, which for example, produce less expensive circuits or are easier to sample from. For instance, these results can be directly transported to the continuous qDrift protocol \cite{Berry2020timedependent}, where the continuous distribution is replaced with an easier one. In~\cite{Berry2020timedependent} the authors already proposed to use a more readily available distribution using norm upperbounds instead of norms. Our results here makes it easier to characterize both the bias and the variance change induced by such a choice. We will see in the next section how to choose $q(j)$ to obtain a guaranteed reduction in the simulation cost. 

Finally, we compute the bound of the expected fluctuations around the true evolution, following the strategy of \cite[Proposition 3.4]{QDRift_caltech}.
\begin{corollary}[Fluctuation bound]
\label{cor_fluctuation}
Let $H$ be a $n$\hyp qubit Hamiltonian, $q(j)$ an arbitrary distribution, $t$ the simulation time, $N$ a fixed number of qDrift samples, and $M$ a fixed number of qDrift experiments. Set $\mathcal{U}_H[\rho]=U_H\rho U^\dagger_H$ (with $U_H = e^{-iHt}$) and take the importance sampled qDrift channel $\mathcal{E}_q(t;N,M)$.
We have 
\begin{equation}
\begin{split}
   \mathbb{E}\left[\left\|\right.\right.&\mathcal{E}_q(t;N,M) -\left.\left.\mathcal{U}_H\right \|_\diamond\right]
    \leq 2\frac{t^2\lambda^2}{N}\left(1+\mathbb{E}_p[\omega]\right)\\
    &+\alpha \frac{nt\lambda}{NM}\left(1+\max_k\omega(k)\right)\\
&+\alpha \sqrt{\frac{n}{NM}}t\lambda\left(1+\max_k\omega(k)\right),
    \end{split}
\end{equation}
with $\alpha$ being a numerical constant depending on $H$.

\end{corollary}

We now have a better understanding of how to choose $N$ and $M$ for a particular distribution $q(j)$ and desired expected accuracy, in diamond norm, $\epsilon$. In particular, if we choose $N$ to control the bias to $\epsilon/\kappa$, with $\kappa>1$, as follows
\begin{equation}
N=2\kappa\frac{t^2\lambda^2}{\epsilon}(1+\mathbb{E}_p[\omega]) \;,
\end{equation}
we then need to choose $M$ so that
\begin{equation}
\alpha t\lambda\left(1+\max_k\omega(k)\right)\left(\frac{n}{NM}+\sqrt{\frac{n}{NM}}\right)\leq\frac{\kappa-1}{\kappa}\epsilon\;.
\end{equation}
Using the choice for $N$ above, we have
\begin{equation}
t\lambda \left(1+\max_{k}\omega(k)\right)=\sqrt{N\frac{\epsilon}{2\kappa}}\frac{1+\max_{k}\omega(k)}{\sqrt{1+\mathbb{E}_p[\omega]}}\;,
\end{equation}
resulting in the condition
\begin{equation}
\alpha\sqrt{\frac{n}{2M}}\frac{1+\max_{k}\omega(k)}{\sqrt{1+\mathbb{E}_p[\omega]}}\left(1+\sqrt{\frac{n}{NM}}\right)\leq(\kappa-1)\sqrt{\frac{\epsilon}{\kappa}}\;.
\end{equation}
Since $N\geq1$, for $M\geq n$ we obtain
\begin{equation}
\label{eq:Mexpect}
M=\frac{n}{\epsilon}\frac{2\alpha^2\kappa}{(\kappa-1)^2}\frac{\left(1+\max_{k}\omega(k)\right)^2}{1+\mathbb{E}_p[\omega]}\;.
\end{equation}
The parameter $\kappa$ can be chosen to reduce $N$ as much as possible while controlling that $M$ does not diverge. In general, for a given $\kappa$ and fixed choice for the distribution $q(j)$, we have the following asymptotic scaling
\begin{equation}
N=\mathcal{O}\left(\frac{t^2\lambda^2}{\epsilon}\right)\, ,\quad M=\mathcal{O}\left(\frac{n}{\epsilon}\right)\;.
\end{equation}
We then see that, thanks to the concentration bound in Theorem~\ref{th:3} depending on $NM$, the number of qDrift experiments scales better than $\mathcal{O}(1/\epsilon^2)$ that one would naively expect from shot noise.

\subsection{Composite channels} 
\label{sec_composite}
We recall that the main goal of this paper is to reduce the actual implementation cost of random product formulas when running on quantum hardware. We remark that our framework can be naturally embedded in the context of composite channels~\cite{Hybrid_Wiebe,hybrid_jin,hybridizedmethodsRajput2022}, and use the deterministic nature of Trotter product formula to further reduce the cost. Following~\cite{Hybrid_Wiebe}, we will first introduce composite channels, unify them with our importance sampling scheme, and compute the relevant bounds on the bias, variance and fluctuation.

Given a partition of the Hamiltonian, a composite channel is a composition of channels, which are used to simulate the different terms in the partition. For simplicity, we will only consider the case where H is split into two parts $H=A+B$ with decompositions
\begin{equation}
\label{eq:ab_decomp}
A=\sum_{i\in I^A}a_iA_i\quad B=\sum_{i\in I^B}b_iB_i\;,
\end{equation}
with $\|A_i\|=\|B_i\|=1$ and $I^A$ and $I^B$ two sets of indices, and use a deterministic formula for $A$ and a stochastic one for $B$. To better understand this paradigm, we first perform an outer first order Trotter decomposition

\begin{equation}
\label{eq:outer_breakup}
e^{itH}\rho e^{-itH} \coloneqq e^{itA}e^{itB}\rho e^{-itB}e^{-itA}+E_{A,B}(t),
\end{equation}

where $E_{A,B}$ is the error term. We then take $\widetilde{\mathcal{U}}_A(t)$ to be an approximation to the unitary channel $\mathcal{U}_A(t)[\rho]=e^{itA}\rho e^{-itA}$ performing the evolution under $A$ (we consider here a first order product formula) and $\mathcal{E}^B_q(t;N,M)$ the importance sampled qDrift channel for the $B$ term. If we define the composite channel as ${\Lambda}_q(t;N,M)=\widetilde{\mathcal{U}}_A(t)\circ \mathcal{E}^B_q(t;N,M)$, and its corresponding average channel $\overline{\Lambda}_q(t;N)$, one can show that its diamond norm distance from the ideal channel $\mathcal{U}_H$ can be bounded as follows (cf.~\cite{Hybrid_Wiebe})

\begin{widetext}
\begin{equation}
\label{error_composite_channel}
\begin{split}
\epsilon \coloneqq& \left\|\mathcal{U}_H(t)-\overline{\Lambda}_q(t;N_B)\right\|_\diamond \\
\leq&\left\|\mathcal{U}_A(t)-\widetilde{\mathcal{U}}_A(t)\right\|_\diamond+\left\|\mathcal{U}_B(t)-\overline{\mathcal{E}}^B_q(t;N_B)\right\|_\diamond+\|E_{A,B}(t)\|_\diamond\\
\leq& t^2\left(\sum_{i<j}a_ia_j\|[A_i,A_j]\|+\frac{1}{2}\sum_{ij}a_ib_j\|[A_i,B_j]\|+\frac{\lambda_B^2 (1+\mathbb{E}_p[\omega(j)])}{N_B} \right). 
\end{split}
\end{equation}
 
If we split the total time into $r$ segments and we use the union bound as usual, we find
\begin{equation}
\begin{split}
\label{eq:cc_bias}
\left\|\mathcal{U}_H(t)-\overline{\Lambda}_q\left(\frac{t}{r};N_B\right)^{\circ r}\right\|_\diamond\leq& \frac{t^2}{r}\left(\sum_{i<j}a_ia_j\|[A_i,A_j]\|+\frac{1}{2}\sum_{ij}a_ib_j\|[A_i,B_j]\|+\frac{\lambda_B^2(1+\mathbb{E}_p[\omega(j)])}{N_B} \right)\;.
\end{split}
\end{equation}
Apart from the use of a general importance sample qDrift, and the use of the improved bound Eq.~\eqref{eq:qdrift_ebound} obtained in~\cite{QDRift_caltech} for the error in the qDrift channel, this is the same result obtain already in~\cite{Hybrid_Wiebe}. Where we depart from their scheme is in the fact that we consider the possibility that evolution under different terms in the expansion could have a different gate cost. If we denote by $C^A_j$ and $C^B_j$ the cost of the term $j$ in either $A$ or $B$, we can bound the total cost of the composite channel as follows
\begin{equation}
\begin{split}
C(\epsilon, t) \leq& r \left(\sum_{l \in I^A} C^A_l + N_B\max_{j\in B}\left(C^B_j\right)\right) \\ 
\leq & \left(\sum_{l \in I^A} C^A_l + N_B\max_{j \in B}\left(C^B_j\right)\right) \frac{t^2}{\epsilon}\left(\sum_{i<j}a_ia_j\|[A_i,A_j]\|+\frac{1}{2}\sum_{ij}a_ib_j\|[A_i,B_j]\|+\frac{\lambda_B^2(1+\mathbb{E}_p[\omega(j)])}{N_B}\right)\\
\end{split}
\end{equation}
The average cost instead can be estimated as 
\begin{equation}
\label{eq:cc_cost}
\begin{split}
\mathbb{E}_q[C(\epsilon,t)]
 \leq &  \left(C^A_{tot} + N_B \mathbb{E}_q\left[C^B\right]\right) \frac{t^2}{\epsilon}\left(\Gamma^{A,B}_{\text{comm}}+\frac{\lambda_B^2 \left(1+\mathbb{E}_p[\omega(j)]\right)}{N_B}\right)\;,
\end{split}
\end{equation}
where we have denoted by $C^A_{tot}=\sum_{l \in A} C^A_l$ the total cost for $A$ and by $\Gamma^{A,B}_{\text{comm}}$ the contribution containing the commutators among the $A$ terms and between the $A$ and $B$ partitions.
\end{widetext}
As was pointed out already in~\cite{Hybrid_Wiebe}, the number of qDrift samples $N_B$ per segment is now a completely free parameter. Following the same strategy adopted there, namely finding explicitly the minimum of the cost, one obtains an optimal number of samples as
\begin{equation}
\label{eq:optN_cc}
N_B=\lambda_B\sqrt{\frac{1+\mathbb{E}_p[\omega(j)]}{\mathbb{E}_q\left[C^B\right]}\frac{C^A_{tot}}{\Gamma^{A,B}_{\text{comm}}}}\;.
\end{equation}
With this choice, the expected cost becomes
\begin{equation}
\begin{split}
\label{eq:opt_cost}
\mathbb{E}_q[C(\epsilon,t)]\leq&\frac{t^2}{\epsilon}\left(\sqrt{\Gamma^{A,B}_{\text{comm}}C^A_{tot}}\right.\\
&\left.+\lambda_B\sqrt{\mathbb{E}_q\left[C^B\right](1+\mathbb{E}_p[\omega(j)])}\right)^2\;.
\end{split}
\end{equation}

Unfortunately, we cannot use directly the result for the concentration bound in Theorem~\ref{th:3} since $r$ compositions of the channel ${\Lambda}_q(t;N,M)$ will require $M^r$ experiments to be implemented. It is thus convenient to introduce another channel defined as
\begin{equation}
\begin{split}
\label{eq:omega_ch}
\Omega_q(t;N,M,r)[\rho] =& \frac{1}{M}\sum_{m=1}^M\Lambda^{m}_q\left(\frac{t}{r};N,1\right)^{\circ r}[\rho],\\
\end{split}
\end{equation}
obtained by averaging $M$ channels composed $r$ times each. The superscript $m$ in $\Lambda^{m}_q$ is used to indicate that for every experiment indexed by $m$ the channel uses a different sample of $Nr$ indices. Note that in the limit of infinite experiments $\overline{\Omega}_q(t;N,r)=\overline{\Lambda}_q\left(\frac{t}{r};N_B\right)^{\circ r}$ and we can still use Eq.~\eqref{eq:cc_bias} to control the bias. For the fluctuations around the average, we instead have following theorem.
\begin{theorem}[Concentration bound for composite channels]
\label{th:cb_cc}
Let $\Omega_q(t;N,M,r)$ be a composite channel on $n$ qubits for the Hamiltonian $H=A+B$ employing an approximate unitary $\widetilde{U}_A\approx e^{-iAt/r}$ for the time evolution under the term $A$ and total time $t/r$, a finite importance sampled qDrift channel $\mathcal{E}_q(t;N,1)$ to approximate evolution under $B$ and $r$ steps using unitaries $W_j=e^{-iB_j\tau_j/r}$ with $\tau_j=(tb_j)/(Nq_j)$. Its concentration around its expectation value can be upper bounded $\forall \epsilon \in [0,4t\lambda_B]$ as
\begin{equation}
\begin{split}
    \text{\normalfont{Pr}}&\left[\norm{\frac{1}{M}\!\left[\prod_{s=r}^1\!\left(\widetilde{U}_A\!\prod_{k=N}^1\! W_{\update{j_{k,s}^m}}\right)\right]\! -\! \left(\!\widetilde{U}_A\mathbb{E}\left[W\right]^N\right)^{r}} \!\geq\!\frac{ \epsilon}{2} \right]  \\
   \leq& 2^{n+1} \exp{- \frac{NMr\epsilon^2}{11t^2\lambda_B^2(1+\max_k\omega(k))^2}}\;.
    \end{split}
\end{equation}
In order to guarantee an approximation error $\epsilon/2$ with probability at least $1-\delta$, it is then sufficient to take
\begin{equation}
NMr = 11\frac{t^2\lambda_B^2}{\epsilon^2}\left(1+\max_k\omega(k)\right)^2(n+1)\log\left(\frac{2}{\delta}\right)\;.
\end{equation}
\end{theorem}
Using the new version of the concentration bound Theorem~\ref{th:cb_cc}, we can also provide an estimate for the expected error of the composite channel.
\begin{corollary}[Fluctuation bound for composite channels]
\label{cor_fluctuation_cc}
Let $H=A+B$ be a $n$\hyp qubit Hamiltonian with decomposition as in Eq.~\eqref{eq:ab_decomp}, $q(j)$ an arbitrary distribution, $t$ the simulation time, $N$ a fixed number of qDrift samples, and $M$ a fixed number of qDrift experiments. Take $\mathcal{U}_H(t)[\rho]=U_H(t)\rho U^\dagger_H(t)$ (with $U_H(t) = e^{-iHt}$), $\widetilde{\mathcal{U}}_A(t)$ a first order Trotter approximation of the channel $\mathcal{U}_A(t)$,  $\mathcal{E}^B_q(t;N,M)$ the importance sampled qDrift channel for the $B$ term and $\Omega_q(t;N,M,r)$ the importance sampled composite channel. 
We then have 
\begin{equation}
\begin{split}
   \mathbb{E}\left[\left\|\right.\right.&\Omega_q(t;N,M,r) -\left.\left.\mathcal{U}_H\right \|_\diamond\right]
    \leq \\
    &2\frac{t^2}{r}\left(\Gamma^{A,B}_{\text{comm}}+\frac{\lambda_B^2}{N}\left(1+\mathbb{E}_p[\omega]\right)\right)\\
    &+\alpha \frac{nt\lambda_B}{NMr}\left(1+\max_k\omega(k)\right)\\
&+\alpha \sqrt{\frac{n}{NMr}}t\lambda_B\left(1+\max_k\omega(k)\right)\;,
    \end{split}
\end{equation}
where the parameter
\begin{equation}
\begin{split}
\Gamma^{A,B}_{\text{comm}}=&\sum_{i<j}a_ia_j\|[A_i,A_j]\|\\
&+\frac{1}{2}\sum_{ij}a_ib_j\|[A_i,B_j]\|\;,
\end{split}
\end{equation}
contains the dependence on commutators.

\end{corollary}

If we now want to ensure the expected error to be less then $\epsilon$ we can take a number of steps given by
\begin{equation}
r=2\kappa\frac{t^2}{\epsilon}\left(\Gamma^{A,B}_{\text{comm}}+\frac{\lambda_B^2}{N}\left(1+\mathbb{E}_p[\omega]\right)\right)\;,
\end{equation}
for some $\kappa>1$ together with $N=N_B$ from Eq.~\eqref{eq:optN_cc} and a number of experiments given by
\begin{equation}
\label{eq:m_cc_opt}
M=\mu_q\frac{n}{\epsilon}\frac{2\alpha^2\kappa}{(\kappa-1)^2}\frac{\lambda_B}{\sqrt{C^A_{tot}}}\;,
\end{equation}
where the dependence on $q(j)$ is absorbed in
\begin{equation}
\label{eq:mu_cc_opt}
\mu_q=\frac{\left(1+\max_k\omega(k)\right)^2\sqrt{\mathbb{E}_q\left[C^B\right]/\left(1+\mathbb{E}_p[\omega]\right)}}{\sqrt{\Gamma^{A,B}_{\text{comm}}C^A_{tot}}+\lambda_B\sqrt{\left(1+\mathbb{E}_p[\omega]\right)\mathbb{E}_q\left[C^B\right]}}.
\end{equation}

A similar derivation to the one carried out here could be used for more general cases, discussed already in~\cite{Hybrid_Wiebe}, when one employs a higher\hyp order Trotter formula for the outer breakup in Eq.~\eqref{eq:outer_breakup} or in the simulation of the evolution under $A$ as well as protocols where the latter is implemented using qubitization. Further improvements obtained by using randomization~\cite{QDRift_caltech,Childs2019fasterquantum,Faehrmann2022randomizingmulti} could also be accommodated.

Having unified the composite channel framework from \cite{Hybrid_Wiebe} with the previous result about importance sampling described in this paper, we will now look into more concrete applications.

\section{Applications}
\label{applications}
\subsection{Reduction of the simulation cost}
\label{sec:cost_red}
 The main idea of this paper is to obtain a computational advantage by choosing the probability distribution $q(j)$ in order to reduce the total simulation cost. For instance, we assign a weight $C_j>0$ to each term $H_j$, representing the number of resources required for its simulation, and choose
\begin{equation}
\label{eq:cred}
q_c(j) = \frac{h_j}{C_j\lambda_c},\quad\quad\lambda_c = \sum_l \frac{h_l}{C_l}\;.
\end{equation}
We will denote this sampling strategy with the subscript $c$, while remarking that the standard qDrift protocol is recovered when the cost is constant $C_j=C$. This framework can be subsumed under the umbrella of importance sampling where the goal is a reduction in the integrated computational cost instead of the variance. The choice of the cost $C_j$ is then dictated by the restrictions from the hardware, and is particularly advantageous if the distribution of the cost has a large variance. We first show that the expected cost of one qDrift sample is always lower when sampling from $q_c(j)$ with respect to $p(j)$, using Jensen's inequality. The complete derivation of the following results can be found in Appendix \ref{app:cost}.
\begin{lemma}
[Jensen's inequality~\cite{Jensen}] Let $X$ be an integrable random variable and  $\varphi(x): \mathbb{R} \rightarrow \mathbb{R}$ a convex function. We then have the following inequality:
\begin{equation}
\varphi(\mathbb{E}[X]) \leq \mathbb{E}[\varphi(X)]
\end{equation}
\label{lemma_jensen}
\end{lemma} 

\begin{corollary}
\label{cor:cred}
The expected cost of an importance sampled qDrift channel with $N=1$ samples and $q(j)=q_c(j)$ is always lower than for the standard qDrift protocol, i.e., we have
\begin{equation}
    \mathbb{E}_{q_c}[C]\leq \mathbb{E}_p[C].
\end{equation}
\end{corollary}
Even if, on average, one qDrift sample is cheaper when sampling from this alternative distribution, this alone might not be enough to claim a reduction in the total simulation cost. This is because, due to Theorem~\ref{th_tighter_bound}, the standard qDrift channel needs less samples at fixed accuracy than an importance sampled one. 
However, we can show that the cost reduction holds in general, as formulated in the following theorem.
\begin{theorem}[Cost reduction - pure qDrift]
\label{th:cost}
Let $N_p(\epsilon,t)$ and $N_{q_c}(\epsilon,t)$ be the number of qDrift samples for the two distributions $p(j) = h_j/\lambda$ and $q_{c}(j) = h_j/(\lambda_{c} C_{j})$ for a given target precision $\epsilon$ and propagation time $t$. The expected cost of the important sampled qDrift channel is then always smaller that the cost of the standard one
\begin{equation}
N_{q_c}(\epsilon,t)\mathbb{E}_{q_c}[C] \leq N_p(\epsilon,t) \mathbb{E}_p[C] .
\end{equation}
The number of experiments is instead increased as
\begin{equation}
M_{q_c}(\epsilon)=M_{p}(\epsilon)\frac{(1+\mathbb{E}_p[1/C]\max_j C_j)^2}{1+\mathbb{E}_p[1/C]\mathbb{E}_p[C]}\;,
\end{equation}
and independent on the total evolution time $t$.
\end{theorem}
We therefore see that, for any simulation using pure qDrift, the importance sampling procedure described in this work guarantees a saving in the cost at the price of increasing the number of independent experiments. A similar saving can also be shown to hold when employing composite channels.

\begin{theorem}[Cost reduction - composite channel]
\label{th:cost_cc}
Let $C_p(\epsilon,t)$ and $C_{q_c}(\epsilon,t)$ be the expected cost to implement the composite channels $\Omega_p(t;N,M,r)$ and $\Omega_{q_c}(t;N,M,r)$ using two distributions $p(j) = h_j/\lambda$ and $q_{c}(j) = h_j/(\lambda_{c} C_{j})$ for a given target precision $\epsilon$ and propagation time $t$. Then the following holds
\begin{equation}
C_{q_c}(\epsilon,t)\leq C_p(\epsilon,t)\;.
\end{equation}
The number of experiments is instead increased, $M_{q_c}(\epsilon)\geq M_{p}(\epsilon)$, but retaining the same scaling with error $\epsilon$ and system size $n$ and also independent on $t$.
\end{theorem}

We note that the specific definition of the cost is specific to each application and hardware. For example, on NISQ devices, the cost is dominated by entangling two\hyp qubit gates, and we might then take their number to define the cost $C$. This choice automatically considers the structure of the device, such as its connectivity and the particular application. For example, when simulating physical systems in second quantization, the importance sampling scheme proposed here will give a lower probability to terms with large Jordan\hyp Wigner strings~\cite{JW}. On the other hand, for applications involving error\hyp corrected devices, one would be tempted to choose the number of $T$ gate instead to define the cost $C$, as they typically take time to be fabricated and are the main bottleneck in the fault\hyp tolerant regime \cite{t_gate_bottleneck}. The situation is, however, less straightforward in this case since the decomposition in terms of $T$\hyp gates can depend, in general, on the choice of sampling probability $q(j)$. For our cost reduction scheme to work, we have made the common assumption that the time evolution under the individual terms $H_j$ can be fast-forwarded; a typical example is when $H_j$ are tensor products of Pauli operators. In these cases, the $T$ cost is directly associated with the implementation of a single qubit z\hyp rotation with an angle determined by the time step $\tau_j$ and therefore on $q(j)$ itself (cf. Eq.~\eqref{eq:tauj}), making it difficult to obtain a good candidate for the coefficients $C_j$. One possibility would be to empirically optimize the sampling distribution $q(j)$, for example, using a Monte Carlo approach or genetic optimization~\cite{genetic}, in order to obtain as many time steps $\tau_j$ as possible equal to integer multiples of $\pi/8$ while preventing the average weight $\mathbb{E}_p[\omega(j)]$ to grow too much. There is, however, a different setup where the importance sampling strategy from Eq.~\eqref{eq:cred} could be employed directly to reduce the overall $T$ count. One can take a decomposition where the terms $H_j$ forming the Hamiltonian are still fast]\hyp forwardable but allow a decomposition in Clifford+T gates whose complexity does not depend on the time (see e.g.~\cite{Gu2021fastforwarding} for possible candidates). In this case, sampling from $q_c(j)$ will also guarantee a cost reduction.

\subsection{Choice of the partitioning}
\label{sssec:tgates}
Composite channels offer great versatility by combining deterministic and random product formulas, but their performance greatly depends on the adopted scheme to partition the Hamiltonian. By inspecting the optimized expression of the expected cost in Eq.~\eqref{eq:opt_cost}, one can already see that a composite channel might be helpful in situations where $\lambda_B$ is very small (as already noticed in~\cite{Hybrid_Wiebe}). For instance, even in the particular case where the terms in $B$ all commute with each other, if $\lambda_B$ is sufficiently small, the cost depends directly only on the cost of implementing the terms in $A$ with the same pre\hyp factor, we would have for a direct first-order Trotter simulation. Thanks to our ability to directly take into account the cost of individual terms, we can also see that another use case is whenever $\mathbb{E}_p\left[C^B\right]\ll C^B_{tot}$ which is possible in situations where the cost of individual terms within $B$ have a significant variation. In addition, for these situations, the distribution $q_c(j)$ from Eq.~\eqref{eq:cred} guarantees a further reduction of the expected cost at the expense of an increase in the required number $M$ of experiments that need to be carried out.

These properties frequently arise in simulations of Effective Field Theories, where higher\hyp order corrections to the interaction are suppressed. In this context, a particularly convenient situation is when an accidental symmetry forces some low\hyp energy constants to take unnaturally small values. In nuclear physics, for instance, typical examples of this phenomenon are the $SU(4)$ Wigner symmetry in systems composed by neutrons and protons~\cite{PhysRev.51.106,KAPLAN1996244} and its generalization to $SU(16)$ in the presence of hyperons~\cite{PhysRevD.96.114510}. For simulations of low energy reactions with nuclei, one could then consider the $SU(4)$ symmetric potential in the deterministic part (as worked out e.g., in~\cite{neutrio_nucleus_roggero}) and add symmetry\hyp breaking terms in a stochastic manner using a qDrift channel. Another example would be the inclusion of effective range effects, absent in purely contact interactions, to improve the accuracy in bulk neutron matter~\cite{PhysRevLett.126.132701} or to provide the required stability to medium mass nuclei~\cite{CONTESSI2017839,LU2019134863}. For situations where the physically relevant value of $\lambda_B$ is not sufficiently small to allow for considerable savings in cost, it could also be possible to rely on extrapolation techniques, e.g., eigenvector continuation~\cite{Frame2018}, to study the system at reduced values of $\lambda_B$ followed by an extrapolation.

More generally, one can also employ more direct optimizations of the splitting by attempting to minimize the cost directly, possibly at the same time as the optimization of the weights in the importance sampling distribution to be used for the stochastic portion of the algorithm. To this end, schemes like the probabilistic partitioning scheme introduced in Ref.~\cite{Hybrid_Wiebe} could prove extremely valuable in enabling substantial savings for any given Hamiltonian one is interested in.

\section{Numerical Simulation}
\label{sec_simulation}
This section presents an application of importance sampling and composite channels for the quantum simulation of a model inspired by a pionless lattice effective field theory \cite{eft}, in particular, a simple toy model for a triton introduced in \cite{neutrio_nucleus_roggero}. We take $A=2$ dynamical nucleons together with a static one (infinite mass) fixed on the first site of a $2\times 2$ lattice with periodic boundary conditions. We will consider the static nucleon to be the proton while the two dynamical ones will be neutrons in two different spin states ($N_f=2$). This model is simple enough to be easily simulated, yet contains much of the leading order contributions to the interaction and can thus provide valuable information about light nuclei and their response functions \cite{response}. The model is equivalent to a two dimensional ($d=2$) Hubbard model with a kinetic hopping term
\begin{equation}
H_{kin}=-T\sum_{f=1}^{N_f}\sum_{<i,j>}c^\dagger_{i,f}c_{j,f}\;,
\end{equation}
together with two and three body interactions
\begin{equation}
    \begin{split}
        H_{int} = &u \sum_{i=1}\sum_{f,f'}^{N_f}n_{i,f}n_{i,f'} + v \sum_{i=1}\sum_{f<f'<f''}n_{i,f}n_{i,f'}n_{i,f''}\\
        &+u \sum_{f=1}^{N_f}n_{1,f}+v\sum_{f<f'}^{N_f}n_{1,f}n_{1,f'},
    \end{split}
\end{equation}
with $T$ the hopping coefficient, $u$ the two\hyp body interaction strength and $v$ the three\hyp body one. We recall that the fermionic operator $c_{i,f}$ destroys a particle of the species $f$ on site $i$, $c^\dagger_{i,f}$ is the corresponding creation operator and $n_{i,f} = c^{\dagger}_{i,f}c_{i,f}$ the number operator. The Hamiltonian becomes particularly simple in first quantization because of the small size of the lattice. By using two qubits to encode the position of each nucleon using the following encoding strategy
\begin{equation}
    |1\rangle \equiv |00\rangle ~~ |2\rangle \equiv |01\rangle ~~ |3\rangle \equiv |10\rangle ~~ |4\rangle \equiv |11\rangle,
\end{equation}
this model can be expressed in the Pauli basis as
\begin{equation}
\begin{split}
H &= 8T+\frac{3u}{4} + \frac{v}{16} - 2T \sum_{i=1}^4 X_i \\
&+\frac{v}{16} \left(\sum_{i<j<k} Z_iZ_jZ_k + Z_1Z_4 + Z_2Z_3 \right)  \\
&+  \left(\frac{u}{4} + \frac{v}{16}\right)\cdot \left(\sum_i Z_i +  Z_1Z_2 + Z_1Z_3 \right.\\
& + Z_2Z_4 + Z_3Z_4 + Z_1Z_2Z_3Z_4 \Biggl),
\end{split}
\label{eq:nuclattice}
\end{equation} 
where $X_k,\,Y_k,\,Z_k$ are the corresponding Pauli matrices acting on qubit $k$. More details about the conversion can be found in the original work~\cite{neutrio_nucleus_roggero}.

Instead of using realistic coefficients from experiments, we split the first quantized Hamiltonian into two parts $A$ and $B$, and define our Hamiltonian as
\begin{equation}
    H^{(j)} =a \sum_{i\in I^A} A^{(j)}_i +b \sum_{i\in I^B} B^{(j)}_i ,
\end{equation}
 where the subscript $j$ denotes different splitting strategies and $I^X$ a set of indices for $X$. We choose uniform coefficients inside each term of the partition in order to have a better control over the error bounds, as well as two different Hamiltonian models. The first is defined through the following separate contributions
\begin{equation}
\begin{split}
A^{(0)}& =  \sum_{k=1}^4 X_k +Z_1Z_4+Z_2Z_3 \\ &+\sum_{1\leq i<j<k\leq 4} Z_iZ_jZ_k
\end{split}
\end{equation}
\begin{equation}
\begin{split}
B^{(0)} &= \sum_{k=1}^4 Z_k + Z_1Z_2+Z_1Z_3 \\ &+Z_2Z_4+Z_3Z_4+Z_1Z_2Z_3Z_4\;.
\end{split}
\end{equation}
The second one is instead given as
\begin{equation}
    \begin{split}
A^{(1)} &= \sum_{k=1}^4 X_k+ Z_1+Z_1Z_4 \\ &+Z_2Z_3+Z_2Z_4+Z_1Z_2Z_4
    \end{split}
\end{equation}
\begin{equation} 
\begin{split} 
B^{(1)} &= \sum_{k=2}^4Z_k+Z_1Z_2+Z_1Z_3+Z_3Z_4+Z_1Z_2Z_3\\&+Z_2Z_3Z_4+Z_1Z_3Z_4+Z_1Z_2Z_3Z_4.
    \end{split}
\end{equation}
For simplicity, we will set $a=1$ and express everything, i.e., the coefficient $b$ and simulation time $t$ in units of $a$.

The first Hamiltonian follows the perturbation theoretical approach, where $A$ is the simpler model $H$ in the $u=-4v$ configuration, where most of the coefficients are zero, and $B$ describes a small perturbation making the system more realistic. A deterministic first order Trotter product formula is used to simulate the bulk of the system $A$, while a qDrift channel handles the contribution from the perturbation $B$. 

The second Hamiltonian is chosen instead in order to  minimize the expected cost of each circuit and exemplifies the role of importance sampling. The $A$ part contains the terms that require small resources to be simulated on the quantum devices, i.e., which have a small cost $C$ in terms of number of two\hyp qubits CNOT gates, while $B$ is composed of the most difficult terms, with some easy single\hyp qubit rotations in order to diminish the expected cost.

\setlength{\tabcolsep}{8pt}
\renewcommand{\arraystretch}{1.3}
\begin{table}
\centering
\begin{tabular}{|c|c||c|c|}
\hline
  generator & cost & generator &cost \\
  \hline
$X_k$    &0.1  & $Z_1Z_2$ &6 \\ 
$Z_k$    &0.1  & $Z_3Z_4$ &6\\
$Z_1Z_4$ &2    & $Z_1Z_2Z_3Z_4$ &6\\ 
$Z_2Z_4$ &2    &   $Z_1Z_3Z_4$ & 8\\ 
$Z_2Z_3$ &2   & $Z_1Z_2Z_3$ &8  \\
$Z_1Z_2Z_4$ &4 &  $Z_1Z_3$&10 \\
$Z_2Z_3Z_4$ &4 & &\\
  \hline
\end{tabular}
  
  \caption{Implementation cost for the different generators appearing in the two considered Hamiltonians.}

 \label{tab:cost}
 \end{table}

 \setlength{\tabcolsep}{8pt}
\renewcommand{\arraystretch}{1.3}
\begin{table}
\centering
\begin{tabular}{|c|cc|}
\hline
  &model 0 & model 1 \\
 \hline 
 $\sum_{i\in I^A}C_i$ &28.4&10.5 \\ 
 $\sum_{i \in I^B} C_i$ &30.4&48.3 \\ 
 $\mathbb{E}_{q_c}[\omega^B]$ &15.43&15.02\\
 \hline
  $\mathbb{E}_p[C^B]$ &3.38 &4.83\\ 
   $\mathbb{E}_{q_c}[C^B]$ &0.22&0.32\\ 
   \hline
  $N_p\mathbb{E}_p[C^B] \cdot \frac{\epsilon}{t^2\lambda^2}$ &6.76 &9.66\\ 
 $N_{q_c}\mathbb{E}_{q_c}[C^B] \cdot \frac{\epsilon}{t^2\lambda^2}$ &3.6 &5.15\\
 \hline
 \end{tabular}
 \caption{Expectation value of the cost of simulating the two terms in the partition $A$ and $B$ for the two different models $j=0,\,j=1$, using either first order Trotterization or (importance sampled) qDrift.}
 \label{tab:exp_cost}
 \end{table}

We assume a constant cost of $0.1$ for the one\hyp qubit operations and of one for every CNOT gate appearing in the multi\hyp qubit terms, after transpilation. We will assume a linear connectivity 1423 and neglect any compilation optimization obtained through gate cancellation from neighboring operations for individual sampled circuits. The cost of each generator is displayed in Table \ref{tab:cost}, while the expected cost for the two different systems for Trotter, plain and importance sampled composite channels, can be seen in Table~\ref{tab:exp_cost}. More precisely, we compute the deterministic cost, the expectation value of the cost per step, and the total cost at fixed accuracy $\epsilon$ and time $t$. We can notice that importance sampling is able to diminish the total simulation cost by a factor of two compared to the plain qDrift and that the use of qDrift channel allows a reduction of an order of magnitude in cost compared to Trotterization. 

\begin{figure*}
\begin{center}
\includegraphics[width=0.85\textwidth]{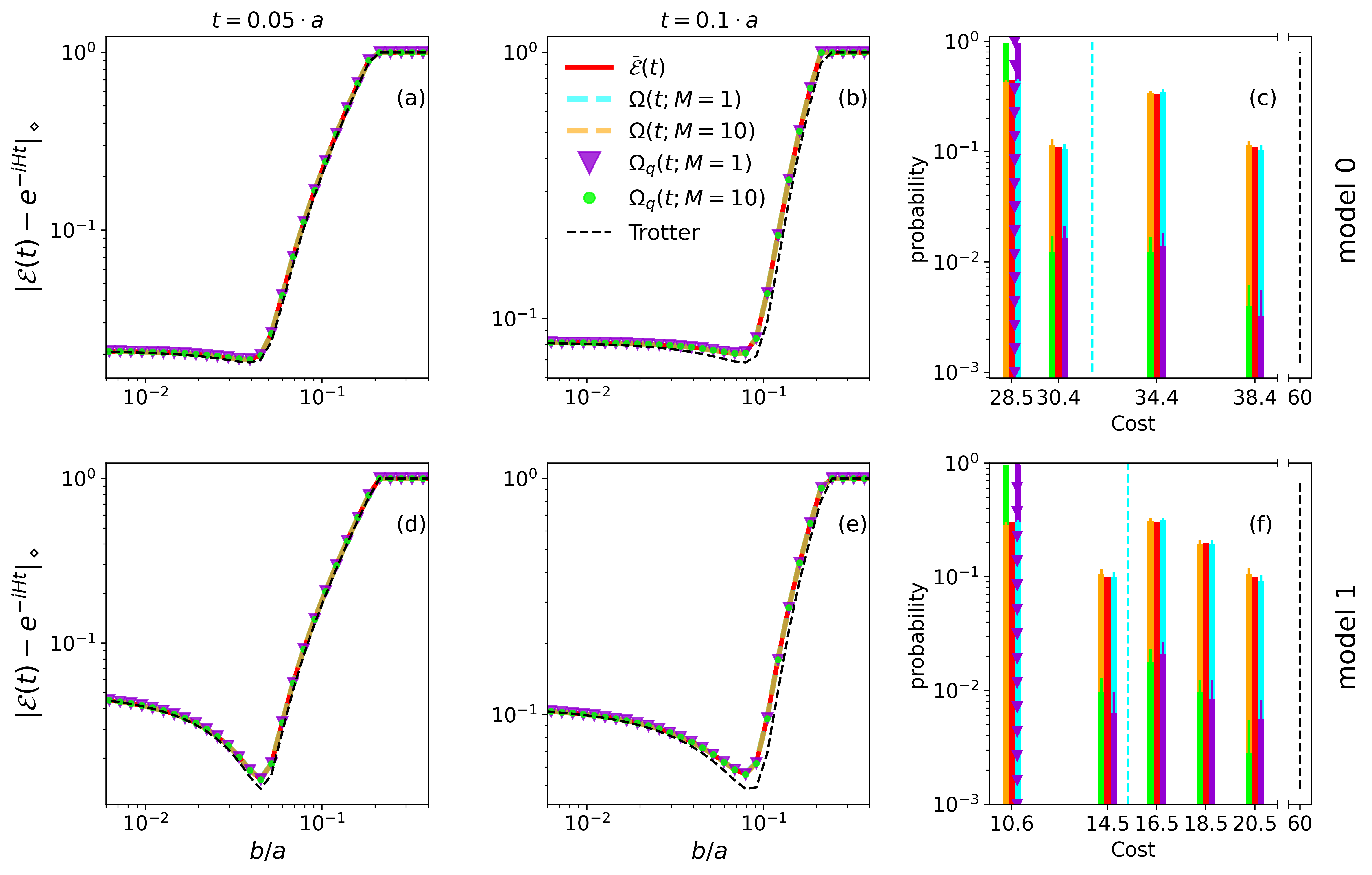}
\caption{Diamond norm [(a), (b), (d), (e)] against the coefficient ratio $b/a$ and histograms of the cost of the quantum circuits [(c), (f)] in terms of CNOT gates for different composite channels (deterministic $\overline{\mathcal{E}}(t)$ in red, standard $\Omega(t; N=1, M, r=1)$ in dashed lines and importance sampled $\Omega_{q}(t; N=1, M, r=1)$ with $q=q_c$ in dots) for the two different models $H^0$ (upper line) and $H^1$ (bottom line), and simulation times ($t=0.05a$ for the first column and $t=0.1a$ for the second one). In panel [(c),(f)], the dashed and dotted lines correspond to the expectation value of the standard and importance sampled qDrift channel, respectively, which is the same for $M=1$ and $M=10$.}
\label{fig:res}
\end{center}
\end{figure*}

We perform quantum simulation of both models $j=0$ and $j=1$ on a noiseless simulator for two different simulation times $t=0.05a$ and $t=0.1a$ and different values of the strength coefficient $b \in [0.005,\,0.5]$, using a composite channel $\Omega_q(t;\,N=1,\,M,\,r=1)$. Part $A$ is evolved using one step of the first order Trotter formula, while part $B$ is evolved using a qDrift channel with $N=1$ step. We consider two different sampling distributions: the standard one $q=p$ and $q=q_c$ with the cost defined as above. The diamond norm, see Eq.~\eqref{diamond_norm}, is displayed in Figure \ref{fig:res} against $b/a$, where the top line shows the computations for the model $j=0$ and the bottom one for the model $j=1$. The diamond distance to the ideal simulation is displayed in dashed black for the full Trotterized channel, in red for the deterministic qDrift $\overline{\mathcal{E}}(t;\,N=1)$ channel, in blue (yellow) dashed line for the standard ($q=p$) composite channel with $M=1$ ($M=10$) and in violet (green) dots for the importance sampled ($q=q_c$) composite channel with $M=1$ ($M=10$). We perform statistics over $R=50$ channels, but since the standard deviations are of order $10^{-4}$, they are not visible. We observe that the errors of the different composite channels are close to each others and are matching the full Trotter channel, except at $b \approx 0.05$ where Trotterization is slightly better. \update{This means that the different channels have approximately the same precision, and we can now look at cost needed to implement them, see panels (c) and (f) on the last column. We show the histograms of the cost over $R=50$ qDrift channels from the sampled circuits obtained with a (importance sampled) composite qDrift channel with $M=1$ or $M=10$ experiments, while the error bars correspond to a $95\%$ confidence interval obtained over $50$ different cost histograms. For sake of readability, contiguos bars of different colors, which are displayed with a small shift over the cost axis, have the same cost written underneath. We observe additionally, that only a precise number of distinct bars are occupied, which correspond to each possible cost under the selected partition. Moreover, the dashed and dotted lines correspond to the expectation value of the standard and importance sampled qDrift channel, respectively, which is the same for $M=1$ and $M=10$. This emphasizes that the cost of the importance sampled channel is close to the expected cost, while the cost of the standard channel is more heterogeneous. However, the most important figure of merit is the distribution of the cost. Hence, the importance sampled qDrift is mainly composed of low\hyp cost circuits, since the probability of sampling a cheap circuit is high, while the standard qDrift samples terms independently of the cost, leading to a more homogeneous cost distribution.} Most importantly, we obtain a reduction in the required amount of resources of a factor 1.8 (2) for the model $j=0$ with the (importance sampled) composite channel and 3.8 (5) for model $j=1$, without sacrificing precision, compared to the full Trotterization. In fact, the cost for the simulation of the $B$ part alone is reduced of one (two) order of magnitude with the pure (importance sampled) qDrift channel compared to direct first order Trotterization, see also Table \ref{tab:exp_cost}. 

\section{Conclusions}
This work generalizes past results on concentration bounds of random product formulae~\cite{QDRift_caltech}, allowing for the introduction of a generic importance sampling scheme. We provide a rigorous characterization of the protocol's bias and statistical fluctuations, extending previous results on qDrift to quantify the algorithm's sample complexity. In particular, we showed that a qDrift channel concentrates exponentially fast in $NM$ around its expectation value, where $N$ is the number of qDrift samples and $M$ the number of experiments, thus allowing an efficient allocation between quantum resources (controlled by $N$) and classical ones (controlled by $M$). These results allow for parallel controllable simulations of a qDrift channel on multiple quantum devices while keeping each circuit shallow enough to mitigate the noise and run time in the NISQ and fault\hyp tolerant era, respectively. Moreover, by incorporating the individual implementation cost for evolution under each of the Hamiltonian terms in a suitable sampling distribution $q_c(j)$, we show that importance sampling obtains a guaranteed total cost reduction, in terms of hardware native cost such as the number of CNOT gates, leading the way to a more straightforward implementation of qDrift in the near term. Under reasonable assumptions, similar cost savings can also be obtained when one wants to reduce the number of $T$ gates, which is beneficial for error\hyp corrected devices. 

In addition, we extend our result to consider composite channels~\cite{Hybrid_Wiebe}, where the Hamiltonian is partitioned into $A$ and $B$, which are simulated separately with a deterministic method (such as a Trotter\hyp Suzuki product formula) and a qDrift channel respectively. We show that the same importance sampling distribution $q_c(j)$ can also be employed in these cases to reduce the quantum resources required for the implementation. In the typical situation where evolution under different terms in the total Hamiltonian incurs different implementation costs, the explicit inclusion of this information in our construction opens the possibility to improve further the savings that can be achieved by using composite channels by optimizing the partitioning schemes~\cite{Hybrid_Wiebe}. In general, it could be profitable to handle Hamiltonian terms, which are expensive but have small norms in a stochastic way using qDrift. We propose different concrete applications within nuclear physics that may benefit from such an approach from an Effective Field Theory perspective.

Finally, the theoretical results are illustrated through numerical simulations of a simple model of a triton on a ($2\times 2$) lattice in first quantization. We find that a significant cost reduction ($5\times$) can be obtained using composite channels and importance sampling without sacrificing accuracy. The approach is robust for different strength values between the two parts of the partition. Due to the quadratic scaling with simulation time of the qDrift part of the scheme, the protocol is particularly well suited for applications that require relatively short evolution times, e.g., protocols to measure observables by signal processing~\cite{Santagati2018,PhysRevA.101.022328,PRXQuantum.2.020317}.

The example importance sampling strategy described in this work has the advantage of being simple and providing a guaranteed cost reduction, but it might not be optimal for some specific problems. We leave for future work exploration of more direct numerical optimizations of the sampling distribution and the partitioning scheme for constructing composite channels. 
\update{In addition, other variance reduction techniques such as Particle-Filters/Sequential Monte Carlo (see e.g.~\cite{Iba2001,Andrieu2010}) or de-randomization (see e.g.~\cite{Huang2021}) could also possibly be used to improve the efficiency of stochastic quantum algorithms like the one described in this work.}

\begin{acknowledgements}
O.K. and M.G. are supported by CERN through the CERN Quantum Technology Initiative.
A.R. is funded by the European Union under Horizon Europe Program - Grant Agreement 101080086 — NeQST.
Views and opinions expressed are however those of the author(s) only and do not necessarily reflect those of CERN, the European Union or European Climate, Infrastructure and Environment Executive Agency (CINEA). Neither CERN, the European Union nor the granting authority can be held responsible for them.
\end{acknowledgements}

\bibliographystyle{unsrtnat}
\bibliography{bibliography}

\begin{thebibliography}{83}
\providecommand{\natexlab}[1]{#1}
\providecommand{\url}[1]{\texttt{#1}}
\expandafter\ifx\csname urlstyle\endcsname\relax
  \providecommand{\doi}[1]{doi: #1}\else
  \providecommand{\doi}{doi: \begingroup \urlstyle{rm}\Url}\fi

\bibitem[Feynman(1982)]{feynman}
R.P. Feynman.
\newblock Simulating physics with computers.
\newblock \emph{Int J Theor Phys}, 21:\penalty0 467–488, 1982.
\newblock \doi{https://doi.org/10.1007/BF02650179}.

\bibitem[Troyer and Wiese(2005)]{sign_Troyer}
Matthias Troyer and Uwe-Jens Wiese.
\newblock Computational complexity and fundamental limitations to fermionic
  quantum monte carlo simulations.
\newblock \emph{Phys. Rev. Lett.}, 94:\penalty0 170201, May 2005.
\newblock \doi{10.1103/PhysRevLett.94.170201}.
\newblock URL \url{https://link.aps.org/doi/10.1103/PhysRevLett.94.170201}.

\bibitem[Lloyd(1996)]{Lloyd_1996}
Seth Lloyd.
\newblock Universal quantum simulators.
\newblock \emph{Science}, 273\penalty0 (5278):\penalty0 1073--1078, 1996.
\newblock \doi{10.1126/science.273.5278.1073}.
\newblock URL
  \url{https://www.science.org/doi/abs/10.1126/science.273.5278.1073}.

\bibitem[Childs et~al.(2018)Childs, Maslov, Nam, Ross, and
  Su]{science_quantum_sim_cost}
Andrew~M. Childs, Dmitri Maslov, Yunseong Nam, Neil~J. Ross, and Yuan Su.
\newblock Toward the first quantum simulation with quantum speedup.
\newblock \emph{Proceedings of the National Academy of Sciences}, 115\penalty0
  (38):\penalty0 9456--9461, 2018.
\newblock \doi{10.1073/pnas.1801723115}.
\newblock URL \url{https://www.pnas.org/doi/abs/10.1073/pnas.1801723115}.

\bibitem[Dumitrescu et~al.(2018)Dumitrescu, McCaskey, Hagen, Jansen, Morris,
  Papenbrock, Pooser, Dean, and Lougovski]{Dumitrescu_2018}
E.~F. Dumitrescu, A.~J. McCaskey, G.~Hagen, G.~R. Jansen, T.~D. Morris,
  T.~Papenbrock, R.~C. Pooser, D.~J. Dean, and P.~Lougovski.
\newblock Cloud quantum computing of an atomic nucleus.
\newblock \emph{Phys. Rev. Lett.}, 120:\penalty0 210501, May 2018.
\newblock \doi{10.1103/PhysRevLett.120.210501}.
\newblock URL \url{https://link.aps.org/doi/10.1103/PhysRevLett.120.210501}.

\bibitem[Roggero and Carlson(2019)]{roggero2019}
Alessandro Roggero and Joseph Carlson.
\newblock Dynamic linear response quantum algorithm.
\newblock \emph{Phys. Rev. C}, 100:\penalty0 034610, Sep 2019.
\newblock \doi{10.1103/PhysRevC.100.034610}.
\newblock URL \url{https://link.aps.org/doi/10.1103/PhysRevC.100.034610}.

\bibitem[Kiss et~al.(2022)Kiss, Grossi, Lougovski, Sanchez, Vallecorsa, and
  Papenbrock]{PRC_Kiss}
Oriel Kiss, Michele Grossi, Pavel Lougovski, Federico Sanchez, Sofia
  Vallecorsa, and Thomas Papenbrock.
\newblock Quantum computing of the $^{6}\mathrm{Li}$ nucleus via ordered
  unitary coupled clusters.
\newblock \emph{Phys. Rev. C}, 106:\penalty0 034325, Sep 2022.
\newblock \doi{10.1103/PhysRevC.106.034325}.
\newblock URL \url{https://link.aps.org/doi/10.1103/PhysRevC.106.034325}.

\bibitem[Hofstetter and Qin(2018)]{Hofstetter_2018}
W~Hofstetter and T~Qin.
\newblock Quantum simulation of strongly correlated condensed matter systems.
\newblock \emph{Journal of Physics B: Atomic, Molecular and Optical Physics},
  51\penalty0 (8):\penalty0 082001, mar 2018.
\newblock \doi{10.1088/1361-6455/aaa31b}.
\newblock URL \url{https://doi.org/10.1088/1361-6455/aaa31b}.

\bibitem[Keenan et~al.(2022)Keenan, Robertson, Murphy, Zhuk, and
  Goold]{quantum_simulations_XXZ}
Nathan Keenan, Niall Robertson, Tara Murphy, Sergiy Zhuk, and John Goold.
\newblock Evidence of kardar-parisi-zhang scaling on a digital quantum
  simulator.
\newblock \emph{ArXiv e-prints}, 2208.12243, 2022.
\newblock \doi{10.48550/ARXIV.2208.12243}.

\bibitem[Grossi et~al.(2023)Grossi, Kiss, De~Luca, Zollo, Gremese, and
  Mandarino]{lmg_grossi}
Michele Grossi, Oriel Kiss, Francesco De~Luca, Carlo Zollo, Ian Gremese, and
  Antonio Mandarino.
\newblock Finite-size criticality in fully connected spin models on
  superconducting quantum hardware.
\newblock \emph{Phys. Rev. E}, 107:\penalty0 024113, Feb 2023.
\newblock \doi{10.1103/PhysRevE.107.024113}.
\newblock URL \url{https://link.aps.org/doi/10.1103/PhysRevE.107.024113}.

\bibitem[Dupont and Moore(2022)]{PRB_LMG}
Maxime Dupont and Joel~E. Moore.
\newblock Quantum criticality using a superconducting quantum processor.
\newblock \emph{Phys. Rev. B}, 106:\penalty0 L041109, Jul 2022.
\newblock \doi{10.1103/PhysRevB.106.L041109}.
\newblock URL \url{https://link.aps.org/doi/10.1103/PhysRevB.106.L041109}.

\bibitem[Monaco et~al.(2023)Monaco, Kiss, Mandarino, Vallecorsa, and
  Grossi]{Monaco_PRB}
Saverio Monaco, Oriel Kiss, Antonio Mandarino, Sofia Vallecorsa, and Michele
  Grossi.
\newblock Quantum phase detection generalization from marginal quantum neural
  network models.
\newblock \emph{Phys. Rev. B}, 107:\penalty0 L081105, Feb 2023.
\newblock \doi{10.1103/PhysRevB.107.L081105}.
\newblock URL \url{https://link.aps.org/doi/10.1103/PhysRevB.107.L081105}.

\bibitem[Jordan et~al.(2012)Jordan, Lee, and Preskill]{QS_QFT_Preskill}
Stephen~P. Jordan, Keith S.~M. Lee, and John Preskill.
\newblock Quantum algorithms for quantum field theories.
\newblock \emph{Science}, 336\penalty0 (6085):\penalty0 1130--1133, 2012.
\newblock \doi{10.1126/science.1217069}.
\newblock URL \url{https://www.science.org/doi/abs/10.1126/science.1217069}.

\bibitem[Shaw et~al.(2020)Shaw, Lougovski, Stryker, and
  Wiebe]{QS_Schwinger_Lougovski}
Alexander~F. Shaw, Pavel Lougovski, Jesse~R. Stryker, and Nathan Wiebe.
\newblock Quantum {A}lgorithms for {S}imulating the {L}attice {S}chwinger
  {M}odel.
\newblock \emph{{Quantum}}, 4:\penalty0 306, August 2020.
\newblock ISSN 2521-327X.
\newblock \doi{10.22331/q-2020-08-10-306}.
\newblock URL \url{https://doi.org/10.22331/q-2020-08-10-306}.

\bibitem[Klco et~al.(2022)Klco, Roggero, and Savage]{Klco_2022}
Natalie Klco, Alessandro Roggero, and Martin~J Savage.
\newblock Standard model physics and the digital quantum revolution: thoughts
  about the interface.
\newblock \emph{Reports on Progress in Physics}, 85\penalty0 (6):\penalty0
  064301, may 2022.
\newblock \doi{10.1088/1361-6633/ac58a4}.
\newblock URL \url{https://dx.doi.org/10.1088/1361-6633/ac58a4}.

\bibitem[Su et~al.(2021{\natexlab{a}})Su, Huang, and
  Campbell]{Su2021nearlytight}
Yuan Su, Hsin-Yuan Huang, and Earl~T. Campbell.
\newblock Nearly tight {T}rotterization of interacting electrons.
\newblock \emph{{Quantum}}, 5:\penalty0 495, July 2021{\natexlab{a}}.
\newblock ISSN 2521-327X.
\newblock \doi{10.22331/q-2021-07-05-495}.
\newblock URL \url{https://doi.org/10.22331/q-2021-07-05-495}.

\bibitem[Ouyang et~al.(2020)Ouyang, White, and Campbell]{Ouyang2020compilation}
Yingkai Ouyang, David~R. White, and Earl~T. Campbell.
\newblock Compilation by stochastic {H}amiltonian sparsification.
\newblock \emph{{Quantum}}, 4:\penalty0 235, February 2020.
\newblock ISSN 2521-327X.
\newblock \doi{10.22331/q-2020-02-27-235}.
\newblock URL \url{https://doi.org/10.22331/q-2020-02-27-235}.

\bibitem[Martínez-Martínez et~al.(2022)Martínez-Martínez, Yen, and
  Izmaylov]{Martinez_partitioning}
Luis~A. Martínez-Martínez, Tzu-Ching Yen, and Artur~F. Izmaylov.
\newblock Assessment of various hamiltonian partitionings for the electronic
  structure problem on a quantum computer using the trotter approximation.
\newblock \emph{ArXiv e-prints}, 2210.10189, 2022.
\newblock \doi{10.48550/ARXIV.2210.10189}.
\newblock URL \url{https://arxiv.org/abs/2210.10189}.

\bibitem[Delgado et~al.(2022)Delgado, Casares, dos Reis, Zini, Campos,
  Cruz-Hern\'andez, Voigt, Lowe, Jahangiri, Martin-Delgado, Mueller, and
  Arrazola]{Li6_batteries_arrazola}
Alain Delgado, Pablo A.~M. Casares, Roberto dos Reis, Modjtaba~Shokrian Zini,
  Roberto Campos, Norge Cruz-Hern\'andez, Arne-Christian Voigt, Angus Lowe,
  Soran Jahangiri, M.~A. Martin-Delgado, Jonathan~E. Mueller, and Juan~Miguel
  Arrazola.
\newblock Simulating key properties of lithium-ion batteries with a
  fault-tolerant quantum computer.
\newblock \emph{Phys. Rev. A}, 106:\penalty0 032428, Sep 2022.
\newblock \doi{10.1103/PhysRevA.106.032428}.
\newblock URL \url{https://link.aps.org/doi/10.1103/PhysRevA.106.032428}.

\bibitem[Su et~al.(2021{\natexlab{b}})Su, Berry, Wiebe, Rubin, and
  Babbush]{1stq_Babbush}
Yuan Su, Dominic~W. Berry, Nathan Wiebe, Nicholas Rubin, and Ryan Babbush.
\newblock Fault-tolerant quantum simulations of chemistry in first
  quantization.
\newblock \emph{PRX Quantum}, 2:\penalty0 040332, Nov 2021{\natexlab{b}}.
\newblock \doi{10.1103/PRXQuantum.2.040332}.
\newblock URL \url{https://link.aps.org/doi/10.1103/PRXQuantum.2.040332}.

\bibitem[Abrams and Lloyd(1999)]{QPE-Lloyd}
Daniel~S. Abrams and Seth Lloyd.
\newblock Quantum algorithm providing exponential speed increase for finding
  eigenvalues and eigenvectors.
\newblock \emph{Phys. Rev. Lett.}, 83:\penalty0 5162--5165, Dec 1999.
\newblock \doi{10.1103/PhysRevLett.83.5162}.
\newblock URL \url{https://link.aps.org/doi/10.1103/PhysRevLett.83.5162}.

\bibitem[Li et~al.(2019)Li, Liu, Wang, Ashhab, Cui, Chen, Peng, and
  Du]{reaction}
Zhaokai Li, Xiaomei Liu, Hefeng Wang, Sahel Ashhab, Jiangyu Cui, Hongwei Chen,
  Xinhua Peng, and Jiangfeng Du.
\newblock Quantum simulation of resonant transitions for solving the
  eigenproblem of an effective water hamiltonian.
\newblock \emph{Phys. Rev. Lett.}, 122:\penalty0 090504, Mar 2019.
\newblock \doi{10.1103/PhysRevLett.122.090504}.
\newblock URL \url{https://link.aps.org/doi/10.1103/PhysRevLett.122.090504}.

\bibitem[Baroni et~al.(2022)Baroni, Carlson, Gupta, Li, Perdue, and
  Roggero]{two_point_roggero}
A.~Baroni, J.~Carlson, R.~Gupta, Andy C.~Y. Li, G.~N. Perdue, and A.~Roggero.
\newblock Nuclear two point correlation functions on a quantum computer.
\newblock \emph{Phys. Rev. D}, 105:\penalty0 074503, Apr 2022.
\newblock \doi{10.1103/PhysRevD.105.074503}.
\newblock URL \url{https://link.aps.org/doi/10.1103/PhysRevD.105.074503}.

\bibitem[Chiesa et~al.(2019)Chiesa, Tacchino, Grossi, Santini, Tavernelli,
  Gerace, and Carretta]{nature_grossi}
A.~Chiesa, F.~Tacchino, M.~Grossi, P.~Santini, I.~Tavernelli, D.~Gerace, and
  S.~Carretta.
\newblock Quantum hardware simulating four-dimensional inelastic neutron
  scattering.
\newblock \emph{Nat. Phys.}, 15:\penalty0 455--459, 2019.
\newblock \doi{https://doi.org/10.1038/s41567-019-0437-4}.

\bibitem[Hall et~al.(2021)Hall, Roggero, Baroni, and Carlson]{PRD_neutrino}
Benjamin Hall, Alessandro Roggero, Alessandro Baroni, and Joseph Carlson.
\newblock Simulation of collective neutrino oscillations on a quantum computer.
\newblock \emph{Phys. Rev. D}, 104:\penalty0 063009, Sep 2021.
\newblock \doi{10.1103/PhysRevD.104.063009}.
\newblock URL \url{https://link.aps.org/doi/10.1103/PhysRevD.104.063009}.

\bibitem[Amitrano et~al.(2023)Amitrano, Roggero, Luchi, Turro, Vespucci, and
  Pederiva]{neutrino_simulation_amitrano}
Valentina Amitrano, Alessandro Roggero, Piero Luchi, Francesco Turro, Luca
  Vespucci, and Francesco Pederiva.
\newblock Trapped-ion quantum simulation of collective neutrino oscillations.
\newblock \emph{Phys. Rev. D}, 107:\penalty0 023007, Jan 2023.
\newblock \doi{10.1103/PhysRevD.107.023007}.
\newblock URL \url{https://link.aps.org/doi/10.1103/PhysRevD.107.023007}.

\bibitem[Roggero et~al.(2020)Roggero, Li, Carlson, Gupta, and
  Perdue]{neutrio_nucleus_roggero}
Alessandro Roggero, Andy C.~Y. Li, Joseph Carlson, Rajan Gupta, and Gabriel~N.
  Perdue.
\newblock Quantum computing for neutrino-nucleus scattering.
\newblock \emph{Phys. Rev. D}, 101:\penalty0 074038, Apr 2020.
\newblock \doi{10.1103/PhysRevD.101.074038}.
\newblock URL \url{https://link.aps.org/doi/10.1103/PhysRevD.101.074038}.

\bibitem[Du et~al.(2021)Du, Vary, Zhao, and Zuo]{du2021}
Weijie Du, James~P. Vary, Xingbo Zhao, and Wei Zuo.
\newblock Quantum simulation of nuclear inelastic scattering.
\newblock \emph{Phys. Rev. A}, 104:\penalty0 012611, Jul 2021.
\newblock \doi{10.1103/PhysRevA.104.012611}.
\newblock URL \url{https://link.aps.org/doi/10.1103/PhysRevA.104.012611}.

\bibitem[Illa and Savage(2022)]{Illa_2022}
Marc Illa and Martin~J Savage.
\newblock Multi-neutrino entanglement and correlations in dense neutrino
  systems.
\newblock \emph{arXiv preprint arXiv:2210.08656}, 2022.
\newblock \doi{https://doi.org/10.48550/arXiv.2210.08656}.

\bibitem[Suzuki(1991)]{Suzuki}
Masuo Suzuki.
\newblock General theory of fractal path integrals with applications to
  many‐body theories and statistical physics.
\newblock \emph{Journal of Mathematical Physics}, 32\penalty0 (2):\penalty0
  400--407, 1991.
\newblock \doi{10.1063/1.529425}.
\newblock URL \url{https://doi.org/10.1063/1.529425}.

\bibitem[Suzuki(1990)]{SUZUKI1990319}
Masuo Suzuki.
\newblock Fractal decomposition of exponential operators with applications to
  many-body theories and monte carlo simulations.
\newblock \emph{Physics Letters A}, 146\penalty0 (6):\penalty0 319--323, 1990.
\newblock ISSN 0375-9601.
\newblock \doi{https://doi.org/10.1016/0375-9601(90)90962-N}.
\newblock URL
  \url{https://www.sciencedirect.com/science/article/pii/037596019090962N}.

\bibitem[Wiebe et~al.(2010)Wiebe, Berry, Høyer, and
  Sanders]{Wiebe_2010_trotter}
Nathan Wiebe, Dominic Berry, Peter Høyer, and Barry~C Sanders.
\newblock Higher order decompositions of ordered operator exponentials.
\newblock \emph{J. Phys. A: Math. Theor.}, 43\penalty0 (6):\penalty0 065203,
  jan 2010.
\newblock \doi{10.1088/1751-8113/43/6/065203}.
\newblock URL \url{https://dx.doi.org/10.1088/1751-8113/43/6/065203}.

\bibitem[Childs et~al.(2021{\natexlab{a}})Childs, Su, Tran, Wiebe, and
  Zhu]{PhysRevX_high_trotter}
Andrew~M. Childs, Yuan Su, Minh~C. Tran, Nathan Wiebe, and Shuchen Zhu.
\newblock Theory of trotter error with commutator scaling.
\newblock \emph{Phys. Rev. X}, 11:\penalty0 011020, Feb 2021{\natexlab{a}}.
\newblock \doi{10.1103/PhysRevX.11.011020}.
\newblock URL \url{https://link.aps.org/doi/10.1103/PhysRevX.11.011020}.

\bibitem[Zhao et~al.(2022)Zhao, Zhou, Shaw, Li, and Childs]{random_input_Child}
Qi~Zhao, You Zhou, Alexander~F. Shaw, Tongyang Li, and Andrew~M. Childs.
\newblock Hamiltonian simulation with random inputs.
\newblock \emph{Phys. Rev. Lett.}, 129:\penalty0 270502, Dec 2022.
\newblock \doi{10.1103/PhysRevLett.129.270502}.
\newblock URL \url{https://link.aps.org/doi/10.1103/PhysRevLett.129.270502}.

\bibitem[Wecker et~al.(2014)Wecker, Bauer, Clark, Hastings, and
  Troyer]{Troyer_gate_count}
Dave Wecker, Bela Bauer, Bryan~K. Clark, Matthew~B. Hastings, and Matthias
  Troyer.
\newblock Gate-count estimates for performing quantum chemistry on small
  quantum computers.
\newblock \emph{Phys. Rev. A}, 90:\penalty0 022305, Aug 2014.
\newblock \doi{10.1103/PhysRevA.90.022305}.
\newblock URL \url{https://link.aps.org/doi/10.1103/PhysRevA.90.022305}.

\bibitem[Reiher et~al.(2017)Reiher, Wiebe, Svore, Wecker, and
  Troyer]{T_gate_Troyer}
Markus Reiher, Nathan Wiebe, Krysta~M. Svore, Dave Wecker, and Matthias Troyer.
\newblock Elucidating reaction mechanisms on quantum computers.
\newblock \emph{Proceedings of the National Academy of Sciences}, 114\penalty0
  (29):\penalty0 7555--7560, 2017.
\newblock \doi{10.1073/pnas.1619152114}.
\newblock URL \url{https://www.pnas.org/doi/abs/10.1073/pnas.1619152114}.

\bibitem[Preskill(2018)]{Preskill2018quantumcomputingin}
John Preskill.
\newblock Quantum {C}omputing in the {NISQ} era and beyond.
\newblock \emph{{Quantum}}, 2:\penalty0 79, August 2018.
\newblock ISSN 2521-327X.
\newblock \doi{10.22331/q-2018-08-06-79}.
\newblock URL \url{https://doi.org/10.22331/q-2018-08-06-79}.

\bibitem[Georgescu et~al.(2014)Georgescu, Ashhab, and Nori]{RevModPhys.86.153}
I.~M. Georgescu, S.~Ashhab, and Franco Nori.
\newblock Quantum simulation.
\newblock \emph{Rev. Mod. Phys.}, 86:\penalty0 153--185, Mar 2014.
\newblock \doi{10.1103/RevModPhys.86.153}.
\newblock URL \url{https://link.aps.org/doi/10.1103/RevModPhys.86.153}.

\bibitem[Tacchino et~al.(2020)Tacchino, Chiesa, Carretta, and
  Gerace]{simulation_tacchino}
Francesco Tacchino, Alessandro Chiesa, Stefano Carretta, and Dario Gerace.
\newblock Quantum computers as universal quantum simulators: State-of-the-art
  and perspectives.
\newblock \emph{Advanced Quantum Technologies}, 3\penalty0 (3):\penalty0
  1900052, 2020.
\newblock \doi{https://doi.org/10.1002/qute.201900052}.

\bibitem[Ferris et~al.(2022)Ferris, Rasmusson, Bronn, and
  Lanes]{quantum_simulations_lanes}
Kaelyn~J. Ferris, A.~J. Rasmusson, Nicholas~T. Bronn, and Olivia Lanes.
\newblock Quantum simulation on noisy superconducting quantum computers.
\newblock \emph{ArXiv e-prints}, 2209.02795, 2022.
\newblock \doi{10.48550/ARXIV.2209.02795}.

\bibitem[Childs et~al.(2019)Childs, Ostrander, and Su]{Childs2019fasterquantum}
Andrew~M. Childs, Aaron Ostrander, and Yuan Su.
\newblock Faster quantum simulation by randomization.
\newblock \emph{{Quantum}}, 3, 2019.
\newblock ISSN 2521-327X.
\newblock \doi{10.22331/q-2019-09-02-182}.
\newblock URL \url{https://doi.org/10.22331/q-2019-09-02-182}.

\bibitem[Faehrmann et~al.(2022)Faehrmann, Steudtner, Kueng, Kieferov{\'{a}},
  and Eisert]{Faehrmann2022randomizingmulti}
Paul~K. Faehrmann, Mark Steudtner, Richard Kueng, M{\'{a}}ria Kieferov{\'{a}},
  and Jens Eisert.
\newblock Randomizing multi-product formulas for {H}amiltonian simulation.
\newblock \emph{{Quantum}}, 6:\penalty0 806, September 2022.
\newblock ISSN 2521-327X.
\newblock \doi{10.22331/q-2022-09-19-806}.
\newblock URL \url{https://doi.org/10.22331/q-2022-09-19-806}.

\bibitem[Wan et~al.(2022)Wan, Berta, and Campbell]{random_QPE}
Kianna Wan, Mario Berta, and Earl~T. Campbell.
\newblock Randomized quantum algorithm for statistical phase estimation.
\newblock \emph{Phys. Rev. Lett.}, 129:\penalty0 030503, Jul 2022.
\newblock \doi{10.1103/PhysRevLett.129.030503}.
\newblock URL \url{https://link.aps.org/doi/10.1103/PhysRevLett.129.030503}.

\bibitem[Cho et~al.(2022)Cho, Berry, and Hsieh]{doubling_random_berry}
Chien~Hung Cho, Dominic~W. Berry, and Min-Hsiu Hsieh.
\newblock Doubling the order of approximation via the randomized product
  formula.
\newblock \emph{ArXiv e-prints}, 2210.11281, 2022.
\newblock \doi{10.48550/ARXIV.2210.11281}.
\newblock URL \url{https://arxiv.org/abs/2210.11281}.

\bibitem[Knee and Munro(2015)]{PhysRevA_random_knee}
George~C. Knee and William~J. Munro.
\newblock Optimal trotterization in universal quantum simulators under faulty
  control.
\newblock \emph{Phys. Rev. A}, 91:\penalty0 052327, May 2015.
\newblock \doi{10.1103/PhysRevA.91.052327}.
\newblock URL \url{https://link.aps.org/doi/10.1103/PhysRevA.91.052327}.

\bibitem[Wallman and Emerson(2016)]{PhysRevA_random_wallman}
Joel~J. Wallman and Joseph Emerson.
\newblock Noise tailoring for scalable quantum computation via randomized
  compiling.
\newblock \emph{Phys. Rev. A}, 94:\penalty0 052325, Nov 2016.
\newblock \doi{10.1103/PhysRevA.94.052325}.
\newblock URL \url{https://link.aps.org/doi/10.1103/PhysRevA.94.052325}.

\bibitem[Poulin et~al.(2011)Poulin, Qarry, Somma, and
  Verstraete]{PhysRevLett_Poulin}
David Poulin, Angie Qarry, Rolando Somma, and Frank Verstraete.
\newblock Quantum simulation of time-dependent hamiltonians and the convenient
  illusion of hilbert space.
\newblock \emph{Phys. Rev. Lett.}, 106:\penalty0 170501, Apr 2011.
\newblock \doi{10.1103/PhysRevLett.106.170501}.
\newblock URL \url{https://link.aps.org/doi/10.1103/PhysRevLett.106.170501}.

\bibitem[Tran et~al.(2021)Tran, Su, Carney, and Taylor]{Tran_2021}
Minh~C. Tran, Yuan Su, Daniel Carney, and Jacob~M. Taylor.
\newblock Faster digital quantum simulation by symmetry protection.
\newblock \emph{PRX Quantum}, 2:\penalty0 010323, Feb 2021.
\newblock \doi{10.1103/PRXQuantum.2.010323}.
\newblock URL \url{https://link.aps.org/doi/10.1103/PRXQuantum.2.010323}.

\bibitem[Campbell(2019)]{QDrift}
Earl Campbell.
\newblock Random compiler for fast hamiltonian simulation.
\newblock \emph{Phys. Rev. Lett.}, 123:\penalty0 070503, Aug 2019.
\newblock \doi{10.1103/PhysRevLett.123.070503}.
\newblock URL \url{https://link.aps.org/doi/10.1103/PhysRevLett.123.070503}.

\bibitem[Chen et~al.(2021)Chen, Huang, Kueng, and Tropp]{QDRift_caltech}
Chi-Fang Chen, Hsin-Yuan Huang, Richard Kueng, and Joel~A. Tropp.
\newblock Concentration for random product formulas.
\newblock \emph{PRX Quantum}, 2:\penalty0 040305, Oct 2021.
\newblock \doi{10.1103/PRXQuantum.2.040305}.
\newblock URL \url{https://link.aps.org/doi/10.1103/PRXQuantum.2.040305}.

\bibitem[Nakaji et~al.(2023)Nakaji, Bagherimehrab, and Aspuru-Guzik]{qSWIFT}
Kouhei Nakaji, Mohsen Bagherimehrab, and Alan Aspuru-Guzik.
\newblock qswift: High-order randomized compiler for hamiltonian simulation.
\newblock \emph{ArXiv e-prints}, 2302.14811, 2023.
\newblock \doi{10.48550/ARXIV.2302.14811}.
\newblock URL \url{https://arxiv.org/abs/2302.14811}.

\bibitem[Berry et~al.(2020)Berry, Childs, Su, Wang, and
  Wiebe]{Berry2020timedependent}
Dominic~W. Berry, Andrew~M. Childs, Yuan Su, Xin Wang, and Nathan Wiebe.
\newblock Time-dependent {H}amiltonian simulation with {$L^1$}-norm scaling.
\newblock \emph{{Quantum}}, 4:\penalty0 254, April 2020.
\newblock ISSN 2521-327X.
\newblock \doi{10.22331/q-2020-04-20-254}.
\newblock URL \url{https://doi.org/10.22331/q-2020-04-20-254}.

\bibitem[Low and Chuang(2017)]{QSP_Chuang}
Guang~Hao Low and Isaac~L. Chuang.
\newblock Optimal hamiltonian simulation by quantum signal processing.
\newblock \emph{Phys. Rev. Lett.}, 118:\penalty0 010501, Jan 2017.
\newblock \doi{10.1103/PhysRevLett.118.010501}.
\newblock URL \url{https://link.aps.org/doi/10.1103/PhysRevLett.118.010501}.

\bibitem[Low and Chuang(2019)]{Low2019hamiltonian}
Guang~Hao Low and Isaac~L. Chuang.
\newblock Hamiltonian {S}imulation by {Q}ubitization.
\newblock \emph{{Quantum}}, 3:\penalty0 163, July 2019.
\newblock ISSN 2521-327X.
\newblock \doi{10.22331/q-2019-07-12-163}.
\newblock URL \url{https://doi.org/10.22331/q-2019-07-12-163}.

\bibitem[Childs and Wiebe(2012)]{Child}
Andrew~M. Childs and Nathan Wiebe.
\newblock Hamiltonian simulation using linear combinations of unitary
  operations.
\newblock \emph{Quantum Information and Computation}, 12\penalty0
  (11\&12):\penalty0 0901--0924, 2012.
\newblock \doi{https://doi.org/10.26421/QIC12.11-12-1}.

\bibitem[Berry and Childs(2012)]{Black_box_Berry}
Dominic~W. Berry and Andrew~M. Childs.
\newblock Black-box hamiltonian simulation and unitary implementation.
\newblock \emph{Quantum Info. Comput.}, 12\penalty0 (1–2):\penalty0 29–62,
  jan 2012.
\newblock ISSN 1533-7146.
\newblock URL \url{https://dl.acm.org/doi/10.5555/2231036.2231040}.

\bibitem[Babbush et~al.(2018)Babbush, Gidney, Berry, Wiebe, McClean, Paler,
  Fowler, and Neven]{PhysRevX.8.041015}
Ryan Babbush, Craig Gidney, Dominic~W. Berry, Nathan Wiebe, Jarrod McClean,
  Alexandru Paler, Austin Fowler, and Hartmut Neven.
\newblock Encoding electronic spectra in quantum circuits with linear t
  complexity.
\newblock \emph{Phys. Rev. X}, 8:\penalty0 041015, Oct 2018.
\newblock \doi{10.1103/PhysRevX.8.041015}.
\newblock URL \url{https://link.aps.org/doi/10.1103/PhysRevX.8.041015}.

\bibitem[Childs et~al.(2021{\natexlab{b}})Childs, Su, Tran, Wiebe, and
  Zhu]{Theory_Trotter}
Andrew~M. Childs, Yuan Su, Minh~C. Tran, Nathan Wiebe, and Shuchen Zhu.
\newblock Theory of trotter error with commutator scaling.
\newblock \emph{Phys. Rev. X}, 11:\penalty0 011020, Feb 2021{\natexlab{b}}.
\newblock \doi{10.1103/PhysRevX.11.011020}.
\newblock URL \url{https://link.aps.org/doi/10.1103/PhysRevX.11.011020}.

\bibitem[Tokdar and Kass(2010)]{importance_sampling_review}
Surya~T. Tokdar and Robert~E. Kass.
\newblock Importance sampling: a review.
\newblock \emph{WIREs Computational Statistics}, 2\penalty0 (1):\penalty0
  54--60, 2010.
\newblock \doi{https://doi.org/10.1002/wics.56}.
\newblock URL
  \url{https://wires.onlinelibrary.wiley.com/doi/abs/10.1002/wics.56}.

\bibitem[Hagan and Wiebe(2022)]{Hybrid_Wiebe}
Matthew Hagan and Nathan Wiebe.
\newblock Composite quantum simulations.
\newblock \emph{ArXiv e-prints}, 2206.06409, 2022.
\newblock \doi{10.48550/ARXIV.2206.06409}.
\newblock URL \url{https://arxiv.org/abs/2206.06409}.

\bibitem[Jin and Li(2021)]{hybrid_jin}
Shi Jin and Xiantao Li.
\newblock A partially random trotter algorithm for quantum hamiltonian
  simulations.
\newblock \emph{ArXiv e-prints}, 2109.07987, 2021.
\newblock \doi{10.48550/ARXIV.2109.07987}.
\newblock URL \url{https://arxiv.org/abs/2109.07987}.

\bibitem[Rajput et~al.(2022)Rajput, Roggero, and
  Wiebe]{hybridizedmethodsRajput2022}
Abhishek Rajput, Alessandro Roggero, and Nathan Wiebe.
\newblock Hybridized {M}ethods for {Q}uantum {S}imulation in the {I}nteraction
  {P}icture.
\newblock \emph{{Quantum}}, 6:\penalty0 780, August 2022.
\newblock ISSN 2521-327X.
\newblock \doi{10.22331/q-2022-08-17-780}.
\newblock URL \url{https://doi.org/10.22331/q-2022-08-17-780}.

\bibitem[Jensen(1906)]{Jensen}
J.~L. W.~V. Jensen.
\newblock Sur les fonctions convexes et les inégalités entre les valeurs
  moyennes.
\newblock \emph{Acta Mathematica}, 30\penalty0 (1):\penalty0 175--193, 1906.
\newblock \doi{10.1007/BF02418571}.

\bibitem[Jordan and Wigner(1928)]{JW}
P.~Jordan and E.~Wigner.
\newblock Über das paulische Äquivalenzverbot.
\newblock \emph{Z. Physik}, 47:\penalty0 631–651, 1928.
\newblock \doi{https://doi.org/10.1007/BF01331938}.

\bibitem[Chamberland and K.(2020)]{t_gate_bottleneck}
C.~Chamberland and Noh K.
\newblock Very low overhead fault-tolerant magic state preparation using
  redundant ancilla encoding and flag qubits.
\newblock \emph{npj Quantum Inf}, 6:\penalty0 91, 2020.
\newblock \doi{https://doi.org/10.1038/s41534-020-00319-5}.

\bibitem[Katoch et~al.(2021)Katoch, Chauhan, and Kumar]{genetic}
Sourabh Katoch, Sumit~Singh Chauhan, and Vijay Kumar.
\newblock A review on genetic algorithm: past, present, and future.
\newblock \emph{Multimed Tools Appl}, 80:\penalty0 8091--8126, 2021.
\newblock \doi{https://doi.org/10.1007/s11042-020-10139-6}.

\bibitem[Gu et~al.(2021)Gu, Somma, and
  {\c{S}}ahino{\u{g}}lu]{Gu2021fastforwarding}
Shouzhen Gu, Rolando~D. Somma, and Burak {\c{S}}ahino{\u{g}}lu.
\newblock Fast-forwarding quantum evolution.
\newblock \emph{{Quantum}}, 5:\penalty0 577, November 2021.
\newblock ISSN 2521-327X.
\newblock \doi{10.22331/q-2021-11-15-577}.
\newblock URL \url{https://doi.org/10.22331/q-2021-11-15-577}.

\bibitem[Wigner(1937)]{PhysRev.51.106}
E.~Wigner.
\newblock On the consequences of the symmetry of the nuclear hamiltonian on the
  spectroscopy of nuclei.
\newblock \emph{Phys. Rev.}, 51:\penalty0 106--119, Jan 1937.
\newblock \doi{10.1103/PhysRev.51.106}.
\newblock URL \url{https://link.aps.org/doi/10.1103/PhysRev.51.106}.

\bibitem[Kaplan and Savage(1996)]{KAPLAN1996244}
David~B. Kaplan and Martin~J. Savage.
\newblock The spin-flavor dependence of nuclear forces from large-n qcd.
\newblock \emph{Physics Letters B}, 365\penalty0 (1):\penalty0 244--251, 1996.
\newblock ISSN 0370-2693.
\newblock \doi{https://doi.org/10.1016/0370-2693(95)01277-X}.
\newblock URL
  \url{https://www.sciencedirect.com/science/article/pii/037026939501277X}.

\bibitem[Wagman et~al.(2017)Wagman, Winter, Chang, Davoudi, Detmold, Orginos,
  Savage, and Shanahan]{PhysRevD.96.114510}
Michael~L. Wagman, Frank Winter, Emmanuel Chang, Zohreh Davoudi, William
  Detmold, Kostas Orginos, Martin~J. Savage, and Phiala~E. Shanahan.
\newblock Baryon-baryon interactions and spin-flavor symmetry from lattice
  quantum chromodynamics.
\newblock \emph{Phys. Rev. D}, 96:\penalty0 114510, Dec 2017.
\newblock \doi{10.1103/PhysRevD.96.114510}.
\newblock URL \url{https://link.aps.org/doi/10.1103/PhysRevD.96.114510}.

\bibitem[Alexandru et~al.(2021)Alexandru, Bedaque, Berkowitz, and
  Warrington]{PhysRevLett.126.132701}
Andrei Alexandru, Paulo Bedaque, Evan Berkowitz, and Neill~C. Warrington.
\newblock Structure factors of neutron matter at finite temperature.
\newblock \emph{Phys. Rev. Lett.}, 126:\penalty0 132701, Apr 2021.
\newblock \doi{10.1103/PhysRevLett.126.132701}.
\newblock URL \url{https://link.aps.org/doi/10.1103/PhysRevLett.126.132701}.

\bibitem[Contessi et~al.(2017)Contessi, Lovato, Pederiva, Roggero, Kirscher,
  and {van Kolck}]{CONTESSI2017839}
L.~Contessi, A.~Lovato, F.~Pederiva, A.~Roggero, J.~Kirscher, and U.~{van
  Kolck}.
\newblock Ground-state properties of 4he and 16o extrapolated from lattice qcd
  with pionless eft.
\newblock \emph{Physics Letters B}, 772:\penalty0 839--848, 2017.
\newblock ISSN 0370-2693.
\newblock \doi{https://doi.org/10.1016/j.physletb.2017.07.048}.
\newblock URL
  \url{https://www.sciencedirect.com/science/article/pii/S0370269317306044}.

\bibitem[Lu et~al.(2019)Lu, Li, Elhatisari, Lee, Epelbaum, and
  Meißner]{LU2019134863}
Bing-Nan Lu, Ning Li, Serdar Elhatisari, Dean Lee, Evgeny Epelbaum, and Ulf-G.
  Meißner.
\newblock Essential elements for nuclear binding.
\newblock \emph{Physics Letters B}, 797:\penalty0 134863, 2019.
\newblock ISSN 0370-2693.
\newblock \doi{https://doi.org/10.1016/j.physletb.2019.134863}.
\newblock URL
  \url{https://www.sciencedirect.com/science/article/pii/S0370269319305775}.

\bibitem[Frame et~al.(2018)Frame, He, Ipsen, Lee, Lee, and Rrapaj]{Frame2018}
Dillon Frame, Rongzheng He, Ilse Ipsen, Daniel Lee, Dean Lee, and Ermal Rrapaj.
\newblock Eigenvector continuation with subspace learning.
\newblock \emph{Phys. Rev. Lett.}, 121:\penalty0 032501, Jul 2018.
\newblock \doi{10.1103/PhysRevLett.121.032501}.
\newblock URL \url{https://link.aps.org/doi/10.1103/PhysRevLett.121.032501}.

\bibitem[Bedaque and van Kolck(2002)]{eft}
Paulo~F. Bedaque and Ubirajara van Kolck.
\newblock Effective field theory for few-nucleon systems.
\newblock \emph{Annual Review of Nuclear and Particle Science}, 52\penalty0
  (1):\penalty0 339--396, 2002.
\newblock \doi{10.1146/annurev.nucl.52.050102.090637}.
\newblock URL \url{https://doi.org/10.1146/annurev.nucl.52.050102.090637}.

\bibitem[Pastore et~al.(2020)Pastore, Carlson, Gandolfi, Schiavilla, and
  Wiringa]{response}
S.~Pastore, J.~Carlson, S.~Gandolfi, R.~Schiavilla, and R.~B. Wiringa.
\newblock Quasielastic lepton scattering and back-to-back nucleons in the
  short-time approximation.
\newblock \emph{Phys. Rev. C}, 101:\penalty0 044612, Apr 2020.
\newblock \doi{10.1103/PhysRevC.101.044612}.
\newblock URL \url{https://link.aps.org/doi/10.1103/PhysRevC.101.044612}.

\bibitem[Santagati et~al.(2018)Santagati, Wang, Gentile, Paesani, Wiebe,
  McClean, Morley-Short, Shadbolt, Bonneau, Silverstone, Tew, Zhou, O’Brien,
  and Thompson]{Santagati2018}
Raffaele Santagati, Jianwei Wang, Antonio~A. Gentile, Stefano Paesani, Nathan
  Wiebe, Jarrod~R. McClean, Sam Morley-Short, Peter~J. Shadbolt, Damien
  Bonneau, Joshua~W. Silverstone, David~P. Tew, Xiaoqi Zhou, Jeremy~L.
  O’Brien, and Mark~G. Thompson.
\newblock Witnessing eigenstates for quantum simulation of hamiltonian spectra.
\newblock \emph{Science Advances}, 4\penalty0 (1):\penalty0 eaap9646, 2018.
\newblock \doi{10.1126/sciadv.aap9646}.
\newblock URL \url{https://www.science.org/doi/abs/10.1126/sciadv.aap9646}.

\bibitem[Roggero and Baroni(2020)]{PhysRevA.101.022328}
A.~Roggero and A.~Baroni.
\newblock Short-depth circuits for efficient expectation-value estimation.
\newblock \emph{Phys. Rev. A}, 101:\penalty0 022328, Feb 2020.
\newblock \doi{10.1103/PhysRevA.101.022328}.
\newblock URL \url{https://link.aps.org/doi/10.1103/PhysRevA.101.022328}.

\bibitem[O'Brien et~al.(2021)O'Brien, Polla, Rubin, Huggins, McArdle, Boixo,
  McClean, and Babbush]{PRXQuantum.2.020317}
Thomas~E. O'Brien, Stefano Polla, Nicholas~C. Rubin, William~J. Huggins, Sam
  McArdle, Sergio Boixo, Jarrod~R. McClean, and Ryan Babbush.
\newblock Error mitigation via verified phase estimation.
\newblock \emph{PRX Quantum}, 2:\penalty0 020317, May 2021.
\newblock \doi{10.1103/PRXQuantum.2.020317}.
\newblock URL \url{https://link.aps.org/doi/10.1103/PRXQuantum.2.020317}.

\bibitem[Iba(2001)]{Iba2001}
Yukito Iba.
\newblock Population monte carlo algorithms.
\newblock \emph{Transactions of the Japanese Society for Artificial
  Intelligence}, 16\penalty0 (2):\penalty0 279--286, 2001.
\newblock \doi{10.1527/tjsai.16.279}.

\bibitem[Andrieu et~al.(2010)Andrieu, Doucet, and Holenstein]{Andrieu2010}
Christophe Andrieu, Arnaud Doucet, and Roman Holenstein.
\newblock Particle markov chain monte carlo methods.
\newblock \emph{Journal of the Royal Statistical Society: Series B (Statistical
  Methodology)}, 72\penalty0 (3):\penalty0 269--342, 2010.
\newblock \doi{https://doi.org/10.1111/j.1467-9868.2009.00736.x}.
\newblock URL
  \url{https://rss.onlinelibrary.wiley.com/doi/abs/10.1111/j.1467-9868.2009.00736.x}.

\bibitem[Huang et~al.(2021)Huang, Kueng, and Preskill]{Huang2021}
Hsin-Yuan Huang, Richard Kueng, and John Preskill.
\newblock Efficient estimation of pauli observables by derandomization.
\newblock \emph{Phys. Rev. Lett.}, 127:\penalty0 030503, Jul 2021.
\newblock \doi{10.1103/PhysRevLett.127.030503}.
\newblock URL \url{https://link.aps.org/doi/10.1103/PhysRevLett.127.030503}.

\bibitem[Tropp(2011)]{martingal}
Joel Tropp.
\newblock {Freedman's inequality for matrix martingales}.
\newblock \emph{Electronic Communications in Probability}, 16\penalty0
  (none):\penalty0 262 -- 270, 2011.
\newblock \doi{10.1214/ECP.v16-1624}.
\newblock URL \url{https://doi.org/10.1214/ECP.v16-1624}.

\end{thebibliography}

\newpage 
\onecolumn
\appendix 
\section{Proofs of the main results}
\label{app:main}
In this section, we present rigorous proofs of the theorems stated in the main results Section \ref{sec_bounds} and Section \ref{sec_composite}. Before being able to provide them, we fist need to state two lemmas from Ref.~\cite{QDRift_caltech}. 
\begin{lemma}
\label{lemma:diamond_bound}
Let $\mathcal{U}(\rho) = U\rho U^\dagger$ and $\mathcal{V}(\rho) = V\rho V^\dagger$  be unitary channels, we then have 
\begin{equation}
\norm{\mathcal{U}-\mathcal{V}}_\diamond \leq 2 \norm{U-V}\;.
\end{equation}
The results carries over to ensembles $(p_k,V_k)$ of unitary channels with weights $p_k\geq0$ for which $\sum_k p_k=1$.
\begin{equation}
\norm{\mathcal{U}-\sum_kp_k\mathcal{V}_k}_\diamond \leq 2 \norm{U-\sum_kp_kV_k}\;.
\end{equation}
\end{lemma}
\begin{lemma}
\label{lemma_exp_bound}
Let $X$ be hermitian. We then have the zero\hyp th order bound $\norm{\exp{iX}-\mathbb{1}}\leq \norm{X}$ and the first\hyp order bound $\norm{\exp{iX}-iX - \mathbb{1}}\leq \frac{1}{2}\norm{X}^2$
\end{lemma}
We are now in a position to prove Theorem \ref{th_tighter_bound}, that we will recall for the ease of the reader. 

\setcounter{theorem}{0}
\begin{theorem}[Bias error bound]

Let $\mathcal{U}(t)$ be a first\hyp order Trotter product channel, $\overline{\mathcal{E}}_q(t;N)$ an average qDrift channel with importance sampling and $\omega(j) = p(j)/q(j)$ the re\hyp weighting factor. The diamond norm distance between these two channels for $N=1$ is then upper bounded by
\begin{equation}
\norm{\mathcal{U}\left(t\right)-\overline{\mathcal{E}}_q(t;1)}_\diamond \leq t^2\lambda^2 \left(1+\mathbb{E}_p\left[\omega(j) \right]\right)\;,
\end{equation}
leading to the following result
\begin{equation}
\norm{\mathcal{U}\left(t\right)-\overline{\mathcal{E}}_q(t;N)}_\diamond \leq \frac{t^2\lambda^2}{N} \left(1+\mathbb{E}_p\left[\omega(j) \right]\right).
\end{equation}
\end{theorem}
\begin{proof}[Proof of Theorem \ref{th_tighter_bound}]
We first note that the Hamiltonian $H$ can be written as the following expectation value
\begin{equation}
    H = \sum_j h_j H_j = \lambda \mathbb{E}_p[H_j] = \lambda \mathbb{E}_q[\omega(j)H_j],
\end{equation}
with $\omega(j)=h_j/(\lambda q(j))$ and therefore
\begin{equation}
\begin{split}
    U\left(t\right) &= e^{-itH} \\
    &= e^{-it\lambda \mathbb{E}_p[H_j]}\\
    &= e^{-it\lambda \mathbb{E}_q[\omega(j)H_j]}\\
    &=e^{-i\mathbb{E}_q[X_j]},
\end{split}
\end{equation}
with $X_j(t) = \frac{t h_j}{ q(j)}H_j$. By noting that
\begin{equation}
\mathbb{E}_q[X_j(t)] \leq \sum_j q(j) X_j(t) = tH\;, 
\end{equation}
we obtain the following bound
\begin{equation}
\left\|\mathbb{E}_q[X_j(t)]\right\| = t\sum_j h_j\|H_j\|= \lambda t\;,
\end{equation}
while
\begin{equation}
\|X_j(t)\| = \left(\frac{th_j}{q(j)}\right)\|H_j\|=\frac{th_j}{q(j)}\;.
\end{equation}

If we denote  $V(t)=\exp{-iX(t)}$ and the corresponding channel as $\mathcal{V}(t)[\rho] = V(t)\rho V(t)^\dagger$, we can express the deterministic qDrift channel with importance sampling as (cf. Eq.~\eqref{qdrift_channel} in the main text)
\begin{equation}
\overline{\mathcal{E}}_q(t;1)[\rho] = \sum_j q(j) V(t)\rho V(t)^\dagger = \mathbb{E}_q[\mathcal{V}(t)]\;.
\end{equation}

We now observe that
\begin{equation}
\begin{split}
\label{eq:bound_eqbar1}
    \norm{\mathcal{U}\left(t\right) - \overline{\mathcal{E}}_q(t;1)}_\diamond=\norm{\mathcal{U}\left(t\right) - \mathbb{E}_q[\mathcal{V}(t)]}_\diamond 
    \leq &2  \norm{U\left(t\right)- \mathbb{E}_q[V(t)]} \\
    =&2\left \| e^{-i\mathbb{E}_q[X(t)]} -\mathbb{1}+ i\mathbb{E}_q[X(t)] +  \mathbb{E}_q\left[\mathbb{1} -iX(t) - e^{-iX(t)}\right]\right\| \\ 
    \leq &2 \norm{e^{-i\mathbb{E}_q[X(t)]} -\mathbb{1}+ i\mathbb{E}_q[X(t)]} 
    + 2\mathbb{E}_q\left[\norm{e^{-iX(t)}-\mathbb{1} +iX(t)}  \right]\\
    \leq &  \norm{\mathbb{E}_q[X(t)]}^2 + \mathbb{E}_q[\norm{X(t)}^2] \\
    \leq & \left(t\lambda\right)^2 +  \mathbb{E}_q\left[\left(\frac{th_j}{q(j)}\right)^2 \right]\\ 
    =&  \left(t\lambda\right)^2\left(1 +  \mathbb{E}_q\left[\omega^2(j) \right]\right) \\ 
    =&  \left(t\lambda\right)^2\left(1 +  \mathbb{E}_p\left[\omega(j) \right]\right) \;,
    \end{split}
\end{equation}

which was the first result what we set out to show. The result for the general average channel with $N>1$ follows by first considering the fact that we can obtain $\overline{\mathcal{E}}_q(t;1)$ by $N$ compositions
\begin{equation}
\overline{\mathcal{E}}_q(t;N) = \overline{\mathcal{E}}_q\left(\frac{t}{N};1\right)\circ\cdots\circ\overline{\mathcal{E}}_q\left(\frac{t}{N};1\right)=\overline{\mathcal{E}}_q\left(\frac{t}{N};1\right)^{\circ N}=\mathbb{E}_q\left[\mathcal{V}\left(\frac{t}{N}\right)\right]^{\circ N}\;.
\end{equation}
Following~\cite{QDRift_caltech} we then decompose the total evolution time $t$ into $N$ steps of duration $t/N$ to find
\begin{equation}
\begin{split}
\norm{\mathcal{U}\left(t\right) - \overline{\mathcal{E}}_q(t;N)}_\diamond =& \norm{\mathcal{U}\left(\frac{t}{N}\right)^{\circ N} - \overline{\mathcal{E}}_q\left(\frac{t}{N};1\right)^{\circ N}}_\diamond\\
=&\norm{\mathcal{U}\left(\frac{t}{N}\right)^{\circ N} - \mathbb{E}_q\left[\mathcal{V}\left(\frac{t}{N}\right)\right]^{\circ N}}_\diamond\\
\leq& 2\left\|U\left(\frac{t}{N}\right)^N-\mathbb{E}_q\left[V\left(\frac{t}{N}\right)\right]^N\right\|\\
\leq&2N\left\|U\left(\frac{t}{N}\right)-\mathbb{E}_q\left[V\left(\frac{t}{N}\right)\right]\right\|\\
\leq&\frac{t^2\lambda^2}{N}\left(1+\mathbb{E}_p[\omega(j)]\right)\;,
\end{split}
\end{equation}
where we used Lemma~\ref{lemma:diamond_bound} for the second line, the union bound for the third and Eq.~\eqref{eq:bound_eqbar1} for the last step.
\end{proof}

Now that we have generalized the bias error bound, we need to understand how a finite importance sampled qDrift channel concentrates around its expected value. This will provide us with an estimate of $M$ and $N$ for a given accuracy $\epsilon$. We will use the martingale formalism and rely on Ref.~\cite{martingal} for a more in\hyp depth consideration. 
\begin{definition}[martingale]
\label{def:mart}
Consider a filtration of the master sigma algebra $\mathcal{F}_0 \subset \mathcal{F}_1 \subset \dots \subset \mathcal{F}$. A \emph{martingale} is a sequence $\{B_0, B_1, \dots\}$ of random variable satisfying 
\begin{enumerate}
    \item $\sigma(B_k) \subset \mathcal{F}_k$ \hfill(causality)
    \item $\mathbb{E}[B_k|B_{k-1}\dots B_0]=B_{k-1}$ \hfill(status quo).
\end{enumerate}
\end{definition}
The intuition one may have is to think of $k$ as a time index and $\mathcal{F}_k$ to contain all events determined by the past up to time $k$. The causality requirement states that the present $B_k$ may only depend on the past $(B_{k-1},\dots,B_0)$, and the status quo conditions formulate that, on average, today is the same as yesterday.

Before proving the theorem, we need to state another result from Ref.~\cite[Corollary 3.4]{QDRift_caltech}. 
\begin{corollary}
\label{corollary_3.4}
Let $\{B_k:k=0,\dots,N\}\subset \mathbb{M}_{dxd}$ be a matrix martingale. Assume that the associated difference $C_k:=B_k-B_{k-1}$ obeys $\norm{C_k}\leq R$ and its conditional variance $\norm{\sum_{k=0}^N\mathbb{E}[C_kC_k^\dagger|C_{k-1}\dots C_0]}\leq v$ almost surely. Then $\forall \tau \geq 0$, we have 
\begin{equation}
\begin{split}
    \text{\normalfont{Pr}}&\left[\norm{B_N-B_0}\leq \tau\right] \geq  2d \exp{\frac{-\tau^2/2}{v + R\tau/3}} \;.
    \end{split}
\end{equation}
\end{corollary}

In order to show Theorem~\ref{th:3}, we will generalize the construction from Ref.~\cite{QDRift_caltech} of a suitable interpolating martingale. Let's start by introducing the unitaries $V_j = e^{-i\tau_j H_j}$, for which $\mathbb{E}_q[V_j]=\mathbb{E}_q[V]$ independently on $j$, 
and consider the situation where we take a set of $M$ separate samples of the $N$ indices forming the product formula resulting in $M$ distinct martingales of the form
\begin{equation}
\label{eq:bmart}
B^m_k = \mathbb{E}_q[V]^{N-k}\prod_{r=k}^1V^m_r\;,
\end{equation}
with $m\in \{1,2,\dots,M\}$ and $V^m_j$ the $j$-th unitary in the $m$-th sample. For technical reasons, we also define, for every value of $k$, $B^0_k=0$ as well as $B^m_k=0$ for all $m>M$.  The causality condition in Definition~\ref{def:mart} is automatically satisfied since $\forall m$ $B_k^m$ is completely determined by the random samples $V_1^m,\dots,V_k^m$ obtained up to the $k$-th step. The second condition can be checked explicitly, in fact
\begin{equation}
\mathbb{E}\left[B_{k+1}^m|B_k^m,\dots,B_0^m\right]=\mathbb{E}_q[V]^{N-k-1}\mathbb{E}_q[V^m_{k+1}]\prod_{r=k}^1V^m_r=B_k^m\;.
\end{equation}
The generalized interpolating martingale needed for our construction can then be defined for $j\in\{0,1,\dots,NM\}$ as follows
\begin{equation}
D_j = \sum_{m=\lfloor j/N\rfloor+2}^MB^{m}_{0} + B^{\lfloor j/N\rfloor+1}_{j\%N}+\sum_{m=1}^{\lfloor j/N\rfloor} B^m_N\;,
\end{equation}
where we denote with $\lfloor a/b\rfloor$ the entire division from $a$ by $b$ and $a\%b$ the rest, i.e., the integer value $a$ modulo $b$. It is straightforward to see that the sum of two independent martingales is also a martingale, making $D_j$ a valid martingale. We have that the first element is given by
\begin{equation}
D_0 = \sum_{m=1}^M B^m_0 = \sum_{m=1}^M \mathbb{E}_q\left[V\right]^N = M\mathbb{E}_q\left[V\right]^N\;,
\end{equation}
while the last element, corresponding to $j=NM$, is given instead by
\begin{equation}
D_{NM} = \sum_{m=1}^M B^m_N = \sum_{m=1}^M\prod_{r=N}^1V^m_r\;.
\end{equation}
For the special case $M=1$ we recover the construction presented in Ref.~\cite{QDRift_caltech}. The elements of the associated difference sequence are given, for $j\in\{1,2,\dots,NM\}$, by
\begin{equation}
\begin{split}
\label{eq:cjseq}
C_j =& D_j-D_{j-1} = B^{\lfloor j/N\rfloor+1}_{j\%N}-B^{\lfloor j/N\rfloor+1}_{j\%N-1}\\
\coloneqq&B^{a_j}_{b_j}-B^{a_j}_{b_j-1}\\
=&\mathbb{E}_q[V]^{N-b_j}\prod_{r=b_j}^1V^{a_j}_r-\mathbb{E}_q[V]^{N-b_j+1}\prod_{r=b_j-1}^1V^{a_j}_r\\
=&\mathbb{E}_q[V]^{N-b_j}\left(V^{a_j}_{b_j}-\mathbb{E}_q[V]\right)\prod_{r=b_j-1}^1V^{a_j}_r\;,
\end{split}
\end{equation}
where, for ease of notation, we have introduced $a_j=\lfloor j/N\rfloor +1$ and $b_j=j\%N$ in the second line. Since both $V_k$ and $\mathbb{E}_q[V]$ are bounded ($\norm{V_k}\leq1$ almost surely and $\norm{\mathbb{E}_q[V]}\leq1$) we can find the following bound
\begin{equation}
\begin{split}
\norm{C_j} =& \norm{V^{a_j}_{b_j}-\mathbb{E}_q[V]}\\
\leq&\norm{e^{-i\tau_{b_j} H_{b_j}}-\mathbb{1}}+\norm{\mathbb{1}-\mathbb{E}_q[e^{-i\tau_k H_k}]}\\
\leq&\norm{e^{-i\tau_{b_j} H_{b_j}}-\mathbb{1}}+\mathbb{E}_q\left[\norm{\mathbb{1}-e^{-i\tau_k H_k}}\right],\\
\end{split}
\end{equation}
where we used the triangle inequality in the second line and Jensen's inequality in the last. Furthermore, since
\begin{equation}
\norm{\tau_kH_k}=\frac{t h_k}{Nq(k)}\leq \frac{t\lambda}{N}\max_k\omega(k)\;,
\end{equation}
almost surely, using Lemma~\ref{lemma_exp_bound}, it also holds almost surely that 
\begin{equation}
\begin{split}
\label{eq:rbound}
\norm{C_j}\leq&\norm{\tau_kH_k}+\mathbb{E}_q\left[\norm{\tau_kH_k}\right]\\
\leq&\frac{t\lambda}{N}\max_k\omega(k)+\mathbb{E}_q[\tau_k]\\
=&\frac{t\lambda}{N}\left(1+\max_k\omega(k)\right)\;.
\end{split}
\end{equation}
Finally, in order to control the variance, we use
\begin{equation}
\begin{split}
\label{eq:vbound}
&\left\|\sum_{j=1}^{NM}\mathbb{E}[C_jC_j^\dagger|C_{j-1}\dots C_0]\right\|\leq NM\max_j\norm{C_j}^2 =M\frac{t^2\lambda^2}{N} \left(1+\max_k\omega(k)\right)^2\;.
\end{split}
\end{equation}
We will now use this construction to show Theorem \ref{th:3}, which is reformulated below.
\begin{theorem}[Concentration bound]
Let $\mathcal{E}_q(t;N,M)$  be a finite importance sampled qDrift channel on $n$ qubits and $V_{\bm{j}_k^m}$ instances of the $NM$ unitaries that compose the channel. such Their concentration around their expectation value can be upper bounded $\forall \epsilon \in [0,4t\lambda]$ as
\begin{equation}
\begin{split}
    &\text{\normalfont{Pr}}\left[\norm{\frac{1}{M}\sum_m^M \prod_{k=N}^1 V_{\bm{j}^m_k} - \mathbb{E}_q\left[V_j\right]^N} \geq \epsilon/2 \right]  
   \leq 2^{n+1} \exp{- \frac{NM\epsilon^2}{11t^2\lambda^2(1+\max_k\omega(k))^2}}\;.
    \end{split}
\end{equation}
In order to guarantee an approximation error $\epsilon/2$ with probability at least $1-\delta$, it is then sufficient to take
\begin{equation}
NM = 11\frac{t^2\lambda^2}{\epsilon^2}\left(1+\max_k\omega(k)\right)^2(n+1)\log\left(\frac{2}{\delta}\right)\;.
\end{equation}
\end{theorem}

\begin{proof}[Proof of Theorem \ref{th:3}]
Using the results in Eq.~\eqref{eq:rbound} and Eq.~\eqref{eq:vbound}, we see that the parameters $R$ and $v$ from Corollary~\ref{corollary_3.4} can be chosen as
\begin{equation}
R=\frac{t\lambda}{N}\left(1+\max_k\omega(k)\right)\text{,}\quad v=MNR^2\;.
\end{equation}
Using Corollary~\ref{corollary_3.4}, we can show that
\begin{equation}
    \begin{split}
   \text{Pr}\left[\norm{\frac{1}{M}\sum_m^M \prod_{k=N}^1 V_{\bm{j}_k^m} - \mathbb{E}_q\left[V_j\right]^N} \geq \tau \right]=&\text{Pr}\left[\norm{\frac{1}{M}\sum_m^M \prod_{k=N}^1 V_{\bm{j}_k^m} - \mathbb{E}_q\left[\frac{1}{M}\sum_m^M \prod_{k=N}^1 V_{\bm{j}_k^m}\right]} \geq \tau \right]\\ 
  = & \text{Pr}\left[\sum_{m=1}^M \prod_{k=N}^1 V_{\bm{j}_k^m} - \mathbb{E}_q\left[\sum_{m=1}^M\prod_{k=N}^1 V_{\bm{j}_k^m}\right] \geq M \tau \right] \\
  =& \text{Pr}\left[\norm{D_{NM}-D_0} \geq M\tau \right] \\
   \leq & 2^{n+1} \exp{-\frac{M^2\tau^2/2 }{v + MR\tau/3}}\\
   =&2^{n+1} \exp{-\frac{3M\tau^2 }{6NR^2 + 2R\tau}}\;.
   \end{split}
\end{equation}
As in Ref.~\cite{QDRift_caltech}, for $\tau\leq NR$ we consider the simpler bound
\begin{equation}
\begin{split}
 \text{Pr}\left[\norm{\frac{1}{M}\sum_m^M \prod_{k=N}^1 V_{\bm{j}_k^m} - \mathbb{E}\left[V\right]^N} \geq \tau \right] 
\leq&2^{n+1} \exp{-\frac{3M\tau^2 }{8NR^2}}\\
=&2^{n+1} \exp{-\frac{3NM\tau^2 }{8(t\lambda\left(1+\max_k\omega(k)\right))^2}}\;.
\end{split}
\end{equation}
A looser sufficient condition $\tau\leq2\lambda t$ can be obtained by noticing that $\max_k\omega(k)\geq 1$, and the equality only holds when all the weights are the same. Substituting $\tau=\epsilon/2$, using $32/3\leq 11$ and Lemma \ref{lemma:diamond_bound}, we obtain the theorem statement.
\end{proof}
Finally, these results can be used to compute the expected fluctuation bound, see Corollary \ref{cor_fluctuation}.
\setcounter{corollary}{0}
\begin{corollary}[Fluctuation bound]
Let $H$ be a $n$\hyp qubit Hamiltonian, $q(j)$ an arbitrary distribution, $t$ the simulation time, $N$ a fixed number of qDrift samples, and $M$ a fixed number of qDrift experiment. Set $\mathcal{U}_H[\rho]=U_H\rho U^\dagger_H$ (with $U_H = e^{-iHt}$) and take  the importance sampled qDrift channel $\mathcal{E}_q(t;N,M)$.
We then have 
\begin{equation}
\begin{split}
   & \mathbb{E}_q\left[\left\|\mathcal{E}_q(t;N,M) -\mathcal{U}_H\right \|_\diamond\right]
    \leq 2\frac{t^2\lambda^2}{N}\left(1+\mathbb{E}_p[\omega]\right)+\alpha \frac{nt\lambda}{NM}\left(1+\max_k\omega(k)\right)
+\alpha \sqrt{\frac{n}{NM}}t\lambda\left(1+\max_k\omega(k)\right).
    \end{split}
\end{equation}

\end{corollary}

\begin{proof}
We prove the corollary by relating the diamond norm distance to the operator norm for ensembles of unitary channels, see Lemma~\ref{lemma:diamond_bound}, and then using the triangle inequality 
\begin{equation}
    \begin{split}
        \mathbb{E}_q\left[\left\|\mathcal{E}_q(t;N,M) -\mathcal{U}_H\right \|_\diamond\right]
         &\leq 2\mathbb{E}_q\left[ \left\|\frac{1}{M}\sum_{m=1}^M \prod_{k=N}^1 V_{\bm{j}_k^m}-U_H\right\| \right]\\
         &\leq  2\left\|U_H-\mathbb{E}_q[V]^N\right\| + 2\mathbb{E}_q\left[\left\| \frac{1}{M}\sum_{m=1}^M \prod_{k=N}^1 V_{\bm{j}_k^m} - \mathbb{E}_q[V]^N \right\| \right]\;.
\end{split}
\end{equation}
The first term can be bounded by using Theorem \ref{th_tighter_bound}, while for the second we have
\begin{equation}
    \begin{split}
       2\mathbb{E}_q\left[\left\| \frac{1}{M}\sum_{m=1}^M \prod_{k=N}^1 V_{\bm{j}_k^m} - \mathbb{E}_q[V]^N \right\|\right]
       = & 2\int_0^\infty \text{Pr}\left(\left\|\frac{1}{M}\sum_{m=1}^M \prod_{k=N}^1 V_{\bm{j}_k^m} - \mathbb{E}_q[V]^N \right\| \geq \tau \right)\,d\tau \\
      \leq & 2\int_0^\infty \min{\left(1,2\cdot 2^{n} e^{-\frac{3M\tau^2}{6NR^2+2R\tau}} \right)}\,d\tau \\
       \leq& \frac{\alpha}{2} \max{\left(\sqrt{\frac{n}{NM}}t\lambda(1+\max_j{\omega(j)}),\frac{nt\lambda (1+\max_j{\omega(j))}}{NM}\right)}\\
       \leq&\alpha\left(\sqrt{\frac{n}{NM}}t\lambda(1+\max_j{\omega(j)})+\frac{nt\lambda (1+\max_j{\omega(j))}}{NM}\right)
    \end{split}
\end{equation}
As in Ref.~\cite{QDRift_caltech}, the integral is evaluated by cutting it into two parts. The first, with a contribution of nearly one, when the denominator in the exponent is bigger than the numerator, i.e., $\tau \leq \max{\left(\sqrt{\frac{2n}{M}}R,\frac{2Rn}{3M}\right)}$, where $\alpha$ suppresses any constant. The contribution for larger $\tau$ is marginal and of order $\mathcal{O}\left(\max{\left(\sqrt{\frac{2}{M}}R,\frac{2R}{3M}\right)}\right)$.
\end{proof}

When employing composite channels, as introduced by Ref.~\cite{Hybrid_Wiebe}, to simulate Hamiltonians formed by two contributions $H=A+B$, we need to generalize the result from Theorem~\ref{th:3} from the channel $\mathcal{E}_q(t;N,M)$ to the following one (cf. Eq.~\eqref{eq:omega_ch} in the main text)
\begin{equation}
\Omega_q(t;N,M,r) = \frac{1}{M}\sum_{m=1}^M\left(\widetilde{\mathcal{U}}_A\left(\frac{t}{r}\right)\circ\mathcal{E}^B_q\left(\frac{t}{r};N,1\right)\right)\circ\cdots\circ\left(\widetilde{\mathcal{U}}_A\left(\frac{t}{r}\right)\circ\mathcal{E}^B_q\left(\frac{t}{r};N,1\right)\right)\;,
\end{equation}
with $r$ outer compositions. Here $\mathcal{E}^B_q(t;N,1)$ is an importance sampled qDrift channel for approximating the evolution under the $B$ term in the Hamiltonian with only one experiment ($M=1$) and $\widetilde{\mathcal{U}}_A(t)$ is a unitary channel that approximates $\mathcal{U}_A(t)[\rho]=e^{-itA}\rho e^{itA}$. In the main text we considered a first-order Trotter approximation but the next theorem applies to more general cases, including $\widetilde{\mathcal{U}}_A$ being an arbitrary unitary matrix.

\begin{theorem}[Concentration bound for composite channels]
Let $\Omega_q(t;N,M,r)$ be a composite channel on $n$ qubits for the Hamiltonian $H=A+B$ employing an approximate unitary $\widetilde{U}_A\approx e^{-iAt/r}$ for the time evolution under the term $A$ and total time $t/r$, a finite importance sampled qDrift channel $\mathcal{E}_q(t;N,1)$ to approximate evolution under $B$ and $r$ steps using unitaries $W_j=e^{-iB_j\tau_j/r}$ with $\tau_j=(tb_j)/(Nq_j)$. Its concentration around its expectation value can be upper bounded $\forall \epsilon \in [0,4t\lambda_B]$ as
\begin{equation}
\begin{split}
    &\text{\normalfont{Pr}}\left[\norm{\frac{1}{M}\left[\prod_{s=r}^1\left(\widetilde{U}_A\prod_{k=N}^1 W_{\bm{j}^m_{ks}}\right)\right] - \left(\widetilde{U}_A\mathbb{E}\left[W\right]^N\right)^{r}} \geq \epsilon/2 \right] 
   \leq 2^{n+1} \exp{- \frac{NMr\epsilon^2}{11t^2\lambda_B^2(1+\max_k\omega(k))^2}}\;.
    \end{split}
\end{equation}
In order to guarantee an approximation error $\epsilon/2$ with probability at least $1-\delta$, it is then sufficient to take
\begin{equation}
NMr = 11\frac{t^2\lambda_B^2}{\epsilon^2}\left(1+\max_k\omega(k)\right)^2(n+1)\log\left(\frac{2}{\delta}\right)\;.
\end{equation}
\end{theorem}
\begin{proof}
The result can be shown by extending the martingales introduced for the proof of the concentration bound Theorem~\ref{th:3}. We start by noticing that $\mathbb{E}_q[W_j]=\mathbb{E}_q[W]$ independent on $j$. For each of the $M$ experiments we consider $r$ sets of $N$ indices for the unitaries forming the product and denote by $W^m_{sj}$ the $j$-th unitary on the $s$-th block for the $m$-th experiment. We can then use them to generalize the martingale from Eq.~\eqref{eq:bmart} to
\begin{equation}
\mathcal{B}_k^m = \left(\prod_{s=r}^{s_k+1}\widetilde{U}_A\mathbb{E}_q[W]^N\right)\left(\widetilde{U}_A\mathbb{E}_q[W]^{N-j_k}\prod_{j=j_k}^1W^m_{s_kj}\right)\left(\prod_{s=s_k-1}^1\widetilde{U}_A\prod_{j=N}^1W^m_{sj}\right)\;,
\end{equation}
for $k=\{0,1,\dots,Nr\}$. Here we have defined the indices as $s_k=\lfloor (k-1)/N \rfloor + 1$, $j_0=0$ while for $k>0$ $j_k=(k-1)\%N+1$. The different definition of $j_0$ is required in order to accommodate the edge case $k=0$. We also require that $\mathcal{B}_k^0=0$ as well as $\mathcal{B}^{m}_k=0$ for $m>M$ as was done before for the $B_k^m$ martingale. The causality condition in Definition~\ref{def:mart} is automatically satisfied since $\forall m$ $\mathcal{B}_k^m$ is completely determined by the random samples $W_{11}^m,\dots,W_{s_kj_k}^m$ obtained up to the $k$-th step. The second condition can be checked explicitly, for $\lfloor (k)/N\rfloor=\lfloor (k-1)/N \rfloor$, implying that $s_{k+1} = s_k$ and $j_{k+1} = j_k + 1$, we have
\begin{equation}
\begin{split}
&\mathbb{E}\left[\mathcal{B}_{k+1}^m|\mathcal{B}_k^m,\dots,\mathcal{B}_0^m\right]  \\ 
=& \left(\prod_{s=r}^{s_k+1}\widetilde{U}_A\mathbb{E}_q[W]^N\right)\left(\widetilde{U}_A\mathbb{E}_q[W]^{N-j_{k+1}}\mathbb{E}_q[W_{s_kj_{k+1}}]\prod_{j=j_{k+1}-1}^1W^m_{s_kj}\right)\left(\prod_{s=s_k-1}^1\widetilde{U}_A\prod_{j=N}^1W^m_{sj}\right) \\
=& \left(\prod_{s=r}^{s_k+1}\widetilde{U}_A\mathbb{E}_q[W]^N\right)\left(\widetilde{U}_A\mathbb{E}_q[W]^{N-j_{k}}\prod_{j=j_{k}}^1W^m_{s_kj}\right)\left(\prod_{s=s_k-1}^1\widetilde{U}_A\prod_{j=N}^1W^m_{sj}\right) \\
= & \mathcal{B}^m_{k}
\end{split}
\end{equation}
while, for $\lfloor k/N\rfloor =\lfloor (k-1)/N\rfloor +1$, i.e., $k=nN$ for some integer $n$ and therefore $s_{k+1}=s_k+1$ together with $j_{k}=N$ and $j_{k+1}=1$, we have the following
\begin{equation}
\begin{split}
&\mathbb{E}\left[\mathcal{B}_{k+1}^m|\mathcal{B}_k^m,\dots,\mathcal{B}_0^m\right] \\
=&\left(\prod_{s=r}^{s_{k+1}+1}\widetilde{U}_A\mathbb{E}_q[W]^N\right)\left(\widetilde{U}_A\mathbb{E}_q[W]^{N-j_{k+1}} \mathbb{E}_q[W_{s_kj_{k+1}}]\prod_{j=j_{k+1}-1}^1W^m_{s_kj}\right) \left(\prod_{s=s_{k+1}-1}^1\widetilde{U}_A\prod_{j=N}^1W^m_{sj}\right)\\ 
=&\left(\prod_{s=r}^{s_{k+1}+1}\widetilde{U}_A\mathbb{E}_q[W]^N\right)\left(\widetilde{U}_A\mathbb{E}[W]^N \right) \left( \widetilde{U}_A\prod_{j=N}^1 W^m_{{(s_{k+1}-1)j}} \right)\left(\prod_{s=s_{k+1}-2}^1\widetilde{U}_A\prod_{j=N}^1W^m_{sj}\right) \\ 
=&\left(\prod_{s=r}^{s_{k}+1}\widetilde{U}_A\mathbb{E}_q[W]^N\right) \left(\widetilde{U}_A\mathbb{E}_q[W]^{N-j_k}\prod_{j=j_k}^1W^m_{s_kj}\right) \left(\prod_{s=s_k-1}^1\widetilde{U}_A\prod_{j=N}^1W^m_{sj}\right) \\ 
=&\mathcal{B}_k^m\;,
\end{split}
\end{equation}
by noting that $N-j_k=0$.

Finally, the new interpolating martingale for the different experiments takes the following form
\begin{equation}
\mathcal{D}_j = \sum_{m=\lfloor j/N \rfloor+2}^M\mathcal{B}^{m}_{0} + \mathcal{B}^{\lfloor j/N \rfloor+1}_{j\%N}+\sum_{m=1}^{\lfloor j/N \rfloor } \mathcal{B}^m_N\;,
\end{equation}
with the special case for $j=0$ and $j=NMr$ given, similarly to before, by
\begin{equation}
\mathcal{D}_0 = \sum_{m=1}^M\mathcal{B}^{m}_{0} = M\left(\prod_{s=r}^{1}\widetilde{U}_A\mathbb{E}_q[W]^N\right)\quad\quad\mathcal{D}_{NMr}=\sum_{m=1}^M\mathcal{B}^{m}_{Nr} =\sum_{m=1}^M\left(\prod_{s=r}^1\widetilde{U}_A\prod_{j=N}^1W^m_{sj}\right)\;.
\end{equation}
Since between consecutive indices the martingale $\mathcal{D}_j$ changes by only a single unitary, the difference sequence $\mathcal{C}_j=\mathcal{D}_j-\mathcal{D}_{j-1}$ have similar properties as $C_j$ in Eq.~\eqref{eq:cjseq} above. In particular
\begin{equation}
\|\mathcal{C}_j\|\leq \left\|\frac{\tau_{k_j}}{r}B_{k_j}\right\| + \mathbb{E}_q\left[\left\|\frac{\tau_k}{r}B_k\right\|\right]\leq \frac{t\lambda_B}{Nr}\left(1+\max_k\omega(k)\right)\;,
\end{equation}
using the same strategy employed to arrive at Eq.~\eqref{eq:rbound}. For the variance instead
\begin{equation}
\left\|\sum_{j=1}^{NMr}\mathbb{E}[\mathcal{C}_j\mathcal{C}_j^\dagger|\mathcal{C}_{j-1}\dots \mathcal{C}_0]\right\|\leq NMr\max_j\norm{\mathcal{C}_j}^2 =M\frac{t^2\lambda_B^2}{Nr} \left(1+\max_k\omega(k)\right)^2
\end{equation}
The result follows then by taking the parameters $R$ and $v$ from Corollary~\ref{corollary_3.4} as
\begin{equation}
R=\frac{t\lambda_B}{Nr}\left(1+\max_k\omega(k)\right)\text{,}\quad v=MNrR^2\;,
\end{equation}
and using Corollary~\ref{corollary_3.4} as was done to show Theorem~\ref{th:3}.
\end{proof}
We are now in a position to show an upperbound for the expected error of a composite channel
\begin{corollary}[Fluctuation bound for composite channels]
Let $H=A+B$ be a $n$\hyp qubit Hamiltonian with decomposition as in Eq.~\eqref{eq:ab_decomp}, $q(j)$ an arbitrary distribution, $t$ the simulation time, $N$ a fixed number of qDrift samples, and $M$ a fixed number of qDrift experiments. Take $\mathcal{U}_H(t)[\rho]=U_H(t)\rho U^\dagger_H(t)$ (with $U_H(t) = e^{-iHt}$), $\widetilde{\mathcal{U}}_A(t)$ a first order Trotter approximation of the channel $\mathcal{U}_A(t)$,  $\mathcal{E}^B_q(t;N,M)$ the importance sampled qDrift channel for the $B$ term and $\Omega_q(t;N,M,r)$ the importance sampled composite channel.
We then have 
\begin{equation}
\begin{split}
   \mathbb{E}\left[\left\|\right.\right.\Omega_q(t;N,M,r) -\left.\left.\mathcal{U}_H\right \|_\diamond\right]
    \leq& 2\frac{t^2}{r}\left(\Gamma^{A,B}_{\text{comm}}+\frac{\lambda_B^2}{N}\left(1+\mathbb{E}_p[\omega]\right)\right)\\
    &+\alpha \frac{nt\lambda_B}{NMr}\left(1+\max_k\omega(k)\right)+\alpha \sqrt{\frac{n}{NMr}}t\lambda_B\left(1+\max_k\omega(k)\right)\;,
    \end{split}
\end{equation}
where the parameter
\begin{equation}
\begin{split}
\Gamma^{A,B}_{\text{comm}}=&\sum_{i<j}a_ia_j\|[A_i,A_j]\|
+\frac{1}{2}\sum_{ij}a_ib_j\|[A_i,B_j]\|\;,
\end{split}
\end{equation}
contains the dependence on commutators.
\end{corollary}
\begin{proof}
We prove the corollary in the same way as we did Corollary~\ref{cor_fluctuation} by relating the diamond norm distance to the operator norm for ensembles of unitary channels, see Lemma~\ref{lemma:diamond_bound}, and then using the triangle inequality 
\begin{equation}
    \begin{split}
        \mathbb{E}\left[\left\|\right.\right.\Omega_q(t;N,M,r) -&\left.\left.\mathcal{U}_H\right \|_\diamond\right]\leq 2\mathbb{E}\left[\left\|\frac{1}{M}\sum_{m=1}^M \prod_{s=r}^1\prod_{k=N}^1 \tilde{U}_AW_{\bm{j}^m_{ks}} -U_H\right \|\right]\\
         &\leq  2\left\|U_H-\left(\widetilde{U}_A\mathbb{E}_q[W]^{N}\right)^r\right\| + 2\mathbb{E}_q\left[\left\| \frac{1}{M}\sum_{m=1}^M \prod_{s=r}^1\prod_{k=N}^1 \tilde{U}_AW_{\bm{j}^m_{ks}} - \left(\widetilde{U}_A\mathbb{E}_q[W]^{N}\right)^r \right\| \right].
    \end{split}
\end{equation}
The first term can be bounded by using Theorem \ref{th_tighter_bound} and the error bounds for composite channel from Eq.~\eqref{eq:cc_bias} to 
\begin{equation}
2\left\|U_H-\mathbb{E}_q[\left(\widetilde{U}_A\mathbb{E}_q[W]^{N}\right)^r\right\| \leq 2\frac{t^2}{r}\left(\Gamma^{A,B}_{\text{comm}}+\frac{\lambda_B^2}{N}\left(1+\mathbb{E}_p[\omega]\right)\right),
\end{equation}
while for the second we have
\begin{equation}
    \begin{split}
       & \mathbb{E}_q\left\| \frac{1}{M}\sum_{m=1}^M \prod_{s=r}^1\prod_{k=N}^1 \tilde{U}_AW_{\bm{j}^m_{ks}} - \left(\widetilde{U}_A\mathbb{E}_q[W]^{N}\right)^r \right\| \\
       = &2 \int_0^\infty \text{Pr}\left(\left\|\frac{1}{M}\sum_{m=1}^M \prod_{s=r}^1\prod_{k=N}^1 \tilde{U}_AW_{\bm{j}^m_{ks}} - \left(\widetilde{U}_A\mathbb{E}_q[W]^{N}\right)^r\right\| \geq \tau \right)\,d\tau \\
      \leq & 2\int_0^\infty \min{\left(1, 2^{n+1} e^{-\frac{3M\tau^2}{6NrR^2+2R\tau}} \right)}\,d\tau \\
       \leq& \frac{\alpha}{2} \max{\left(\sqrt{\frac{n}{NMr}}t\lambda_B(1+\max_k{\omega(k)}),\frac{nt\lambda_B (1+\max_k{\omega(k)})}{NMr}\right)}\\
       \leq& \alpha \left(\sqrt{\frac{n}{NMr}}t\lambda_B(1+\max_k{\omega(k)})+\frac{nt\lambda_B (1+\max_k{\omega(k)})}{NMr}\right)\;,
    \end{split}
\end{equation}
with $R=\frac{t\lambda_B}{Nr}\left(1+\max_k{\omega(k)}\right)$ from above.
As in Ref.~\cite{QDRift_caltech}, the integral is evaluated by cutting it into two parts. The first, with a contribution of nearly one, when the denominator in the exponent is bigger than the numerator, i.e., $\tau \leq \max{\left(\sqrt{\frac{2n}{M}}R,\frac{2Rn}{3M}\right)}$, where $\alpha$ suppresses any constant. The contribution for larger $\tau$ is marginal and of order $\mathcal{O}\left(\max{\left(\sqrt{\frac{2}{M}}R,\frac{2R}{3M}\right)}\right)$.

\end{proof}

\section{Proofs for the cost reduction}
\label{app:cost}
In this section, we will recall and show the results for the particular distribution $q_c(j)=h_j/C_j$, where $C_j$ is the implementation cost of the corresponding term. 
\begin{corollary}
The expected cost of an important sampled qDrift channel with $N=1$ sample and  $q(j)=q_c(j)$ is always lower than for the standard qDrift
\begin{equation}
    \mathbb{E}_{q_c}[C]\leq \mathbb{E}_p[C].
\end{equation}
\end{corollary}

\begin{proof}[Proof of Corollary \ref{cor:cred}]
\begin{equation}
    \begin{split}
   {\mathbb{E}_{q_c}[C]} =& \sum_{j=1}^{L} q(j) C_j 
    = \frac{\sum_{j=1}^{L}\frac{h_j}{C_j}C_j}{\sum_{j=1}^{L}\frac{h_j}{C_j}}
    = \frac{\sum_{j=1}^{L}h_j}{\sum_{j=1}^{L}\frac{h_j}{C_j}}
     \label{jensen}
    =\frac{1}{\mathbb{E}_p[\frac{1}{C}]}
    \leq \mathbb{E}_p[C],
    \end{split}
\end{equation}
where we used in the last step Jensen's inequality with $\varphi(C)=1/C$, which is convex for real positive numbers $C>0$.
\end{proof} 

This result shows that, on average, unitaries sampled from $q(j)=q_c(j)$ are cheaper to implement than from $q(j)=p(j)$. It remains to be shown if the total implementation cost at fixed accuracy is also reduced with this choice of distribution.
\begin{theorem}[Cost reduction - pure qDrift]
Let $N_p$ and $N_{q_c}$ be the number of qDrift samples for the two distributions $p(j) = h_j/\lambda$ and $q_{c}(j) = h_j/(\lambda_{c} C_{j})$ for a given target precision $\epsilon$. The expected cost of the importance sampled qDrift channel is then always smaller than the standard one
\begin{equation}
N_{q_c}\mathbb{E}_{q_c}[C] \leq N_p \mathbb{E}_p[C] .
\end{equation}
The number of experiments is instead increased as
\begin{equation}
M_{q_c}(\epsilon,t)=M_{p}(\epsilon,t)\frac{(1+\mathbb{E}_p[1/C]\max_j C_j)^2}{1+\mathbb{E}_p[1/C]\mathbb{E}_p[C]}\;
\end{equation}
and, more particularly,  independent on the total evolution time $t$.
\end{theorem}

\begin{proof}[Proof of Theorem \ref{th:cost}]
The factor between $N_p$ and $N_{q_c}$, see Eq.~\eqref{eq_N}, can be expanded as follows
\begin{equation}
\begin{split}
\mathbb{E}_p\left[\omega(j)\right] &= \sum_j\frac{h_j}{\lambda}\omega(j)=\sum_j\lambda_c\frac{h_jC_j}{\lambda^2}\\
&=\mathbb{E}_p[C]\frac{\lambda_c}{\lambda}=\mathbb{E}_p[C]\mathbb{E}_p[1/C].
\end{split}
\end{equation}
Using this result, together with Corollary~\ref{cor:cred} and Theorem~\ref{th_tighter_bound}, we find that a sufficient choice for $N_{q_c}$ to guarantee error $\epsilon$ over a total time $t$ is at least
\begin{equation}
N_{q_c}=\frac{t^2\lambda^2}{\epsilon}\left(1+\mathbb{E}_p[\omega(j)]\right),
\end{equation}
which translates into an average total cost of
\begin{equation}
\begin{split}
C_{q_c}&=\frac{t^2\lambda^2}{\epsilon}\left(1+\mathbb{E}_p[\omega(j)]\right)\mathbb{E}_{q_c}[C]
=\frac{t^2\lambda^2}{\epsilon}\frac{1+\mathbb{E}_p[C]\mathbb{E}_p[1/C]}{\mathbb{E}_p[1/C]}
\end{split}
\end{equation}
On the other hand, using regular qDrift we have
\begin{equation}
C_p=2\frac{t^2\lambda^2}{\epsilon}\mathbb{E}_p[C]\;.
\end{equation}
In order to guarantee a cost reduction we then need
\begin{equation}
\frac{1+\mathbb{E}_p[C]\mathbb{E}_p[1/C]}{\mathbb{E}_p[1/C]}\leq 2\mathbb{E}_p[C]\;,
\end{equation}
or equivalently
\begin{equation}
\mathbb{E}_p[C]\mathbb{E}_p[1/C] \geq 1\;.
\end{equation}
This is always satisfied, as one can easily show using Jensen's inequality.
The result on the increase in the number of experiments follows instead directly from the definition of the distribution $q_c(j)$ and the sufficient condition Eq.~\eqref{eq:Mexpect} derived from Corollary~\ref{cor_fluctuation}.
\end{proof} 

We will finally show that the cost reduction is also retained when considering composite channels.

\begin{theorem}[Cost reduction - composite channel]
Let $C_p(\epsilon,t)$ and $C_{q_c}(\epsilon,t)$ be the expected cost to implement the composite channels $\Omega_p(t;N,M,r)$ and $\Omega_{q_c}(t;N,M,r)$ using two distributions $p(j) = h_j/\lambda$ and $q_{c}(j) = h_j/(\lambda_{c} C_{j})$ for a given target precision $\epsilon$ and propagation time $t$. Then the following holds
\begin{equation}
\mathbb{E}_{q_c}[C(\epsilon,t)]\leq \mathbb{E}_p[C(\epsilon,t)].
\end{equation}
The number of experiments is instead increased, $M_{q_c}(\epsilon)\geq M_{p}(\epsilon)$ 

but retaining the same scaling with error $\epsilon$ and system size $n$ and also independent on the total evolution time $t$.\end{theorem}
\begin{proof}
As from Eq.~\eqref{eq:opt_cost}, we know that the cost of the composite channel can be written as 

\begin{equation}
C(\epsilon,t)=\frac{t^2}{\epsilon}\left(\sqrt{\Gamma^{A,B}_{\text{comm}}C^A_{tot}}+\lambda_B\sqrt{\mathbb{E}_{q_c}\left[C^B\right](1+\mathbb{E}_p[\omega(j)])}\right)^2\;.
\end{equation}
Since the first term is independent from the choice of the sampling distribution, we only need to consider the second one which also appears in Theorem \ref{th:cost} and can therefore be bounded as
\begin{equation}
    \mathbb{E}_{q_c}\left[C^B\right](1+\mathbb{E}_p[\omega(j)])\leq 2\mathbb{E}_p\left[C^B\right]\;.
\end{equation}
For the number of experiments instead, using Eq.~\eqref{eq:m_cc_opt} and Eq.~\eqref{eq:mu_cc_opt}, for the $p(j)$ distribution we have
\begin{equation}
M_p(\epsilon)=\frac{n}{\epsilon}\frac{2\alpha^2\kappa}{(\kappa-1)^2}\frac{\lambda_B}{\sqrt{C^A_{tot}}}
\frac{2^{3/2}\sqrt{\mathbb{E}_p\left[C^B\right]}}{\sqrt{\Gamma^{A,B}_{\text{comm}}C^A_{tot}}+\lambda_B\sqrt{2\mathbb{E}_p\left[C^B\right]}},
\end{equation}
while for the importance sampled distribution $q_c(j)$ we find
\begin{equation}
\begin{split}
M_{q_c}(\epsilon)&=\frac{n}{\epsilon}\frac{2\alpha^2\kappa}{(\kappa-1)^2}\frac{\lambda_B}{\sqrt{C^A_{tot}}}
\frac{\left(1+\mathbb{E}_p[1/C^B]\max_jC^B_j\right)^2\sqrt{\frac{1}{\left(1+\mathbb{E}_p[1/C^B]\mathbb{E}_p[C^B]\right)\mathbb{E}_p[1/C^B]}}
}{\sqrt{\Gamma^{A,B}_{\text{comm}}C^A_{tot}}+\lambda_B\sqrt{\left(1+\mathbb{E}_p[1/C^B]\mathbb{E}_p[C^B]\right)/\mathbb{E}_p[1/C^B]}}\\
&\geq\frac{n}{\epsilon}\frac{2\alpha^2\kappa}{(\kappa-1)^2}\frac{\lambda_B}{\sqrt{C^A_{tot}}}
\frac{\left(1+\mathbb{E}_p[1/C^B]\max_jC^B_j\right)^2\sqrt{\frac{1}{\left(1+\mathbb{E}_p[1/C^B]\mathbb{E}_p[C^B]\right)\mathbb{E}_p[1/C^B]}}
}{\sqrt{\Gamma^{A,B}_{\text{comm}}C^A_{tot}}+\lambda_B\sqrt{2\mathbb{E}_p[C^B]}}\\
&\geq\frac{n}{\epsilon}\frac{2\alpha^2\kappa}{(\kappa-1)^2}\frac{\lambda_B}{\sqrt{C^A_{tot}}}
\frac{2\sqrt{1+\mathbb{E}_p[1/C^B]\max_jC^B_j}\sqrt{\mathbb{E}_p[C^B]}
}{\sqrt{\Gamma^{A,B}_{\text{comm}}C^A_{tot}}+\lambda_B\sqrt{2\mathbb{E}_p[C^B]}}\\
&\geq \frac{n}{\epsilon}\frac{2\alpha^2\kappa}{(\kappa-1)^2}\frac{\lambda_B}{\sqrt{C^A_{tot}}}
\frac{2^{3/2}\sqrt{\mathbb{E}_p\left[C^B\right]}}{\sqrt{\Gamma^{A,B}_{\text{comm}}C^A_{tot}}+\lambda_B\sqrt{2\mathbb{E}_p\left[C^B\right]}}=M_p(\epsilon),
\end{split}
\end{equation}
where the second line is obtained by remarking that, due to Jensen's inequality, 
\begin{equation}
\frac{1+\mathbb{E}_p[C^B]\mathbb{E}_p[1/C^B]}{\mathbb{E}_p[1/C^B]}\leq 2\mathbb{E}_p[C^B]\;.
\end{equation}
In order to get to the third line we used the inequality
\begin{equation}
\frac{\left(1+\mathbb{E}_p[1/C^B]\max_jC^B_j\right)^3}{1+\mathbb{E}_p[1/C^B]\mathbb{E}_p[C^B]}\geq4\mathbb{E}_p[1/C^B]\mathbb{E}_p[C^B]\;,
\end{equation}
and the last inequality follows from $1+\mathbb{E}_p[1/C^B]\max_jC^B_j\geq2$.
\end{proof}

\end{document}